\documentclass[10pt]{article}

\usepackage{amsmath}  
\usepackage{graphicx}
\usepackage{caption}
\usepackage{color}
\usepackage[margin=0.5in]{geometry}
\usepackage{multirow}
\usepackage{tabularx}
\usepackage{cite}
\usepackage[font=footnotesize,labelfont=bf]{caption}
\usepackage{float}
\usepackage{amssymb}
\usepackage{amsmath}
\usepackage{subfig}
\usepackage{multicol}

\title{\bf{Can Wigner distribution functions with collisions satisfy complete positivity and energy conservation?}}
\author{{Matteo Villani}$^{1}$, {Xavier Oriols}$^{1}$}
\date{\small{1: Department of Electronic Engineering, Universitat Aut\`onoma de Barcelona, Campus de la UAB, 08193 Bellaterra, Barcelona, Spain}}

\begin{document}
\maketitle
\begin{abstract}
To avoid the computational burden of many-body quantum simulation, the interaction of an electron with a photon (phonon) is typically accounted for by disregarding the explicit simulation of the photon (phonon) degree of freedom and just modelling its effect on the electron dynamics. For quantum models developed from the (reduced) density matrix or its Wigner-Weyl transformation, the modelling of collisions may violate complete positivity (precluding the typical probabilistic interpretation). In this paper, we show that such quantum transport models can also strongly violate the energy conservation in the electron-photon (electron-phonon) interactions. After comparing collisions models to exact results for an electron interacting with a photon, we conclude that there is no fundamental restriction that prevents a collision model developed within the (reduced) density matrix or Wigner formalisms to satisfy simultaneously complete positivity and energy conservation. However, at the practical level, the development of such satisfactory collision model seems very complicated. Collision models with an explicit knowledge of the microscopic state ascribed to each electron seems recommendable, since they allow to model collisions of each electron individually in a controlled way satisfying both conditions.
\end{abstract}

\paragraph{keywords:}$\!\!$Matter-light interaction; Wigner Function; Complete positivity; Energy conservation; Bohmian conditional wavefunctio
\paragraph{Corresponding author:}$\!\!$xavier.oriols@uab.es
\begin{multicols}{2}
Electron devices are quantum systems outside of thermodynamic equilibrium with many interacting particles (electrons, atoms, photons, etc). In addition, one is not only interested on the time-independent or steady state (DC) simulation of such devices, but also on their time-dependent (AC, transients) performance,  and even on their noise properties. As a result, from a computational point of view, an electron devices is one of the most difficult quantum systems to be simulated. Any attempt to directly get the device performance from the simulation of \textit{all} particles, in a type of solution of the many-body Schr\"odinger equation, is directly impossible. This difficulty has the origin in the so-called many body problem \cite{open}. The typical strategy to reduce such inaccessible computational burden is introducing an artificial division between the simulated particles, usually named the \textit{open system}, and the rest of non-simulated particles, which are usually referred as the \textit{environment} \cite{open,vega}. Such division, in turn, requires introducing the effect of the non-simulated (\textit{environment}) particles onto the simulated ones through some \textit{new} term in the equations of motion of simulated particles. This \textit{new} element in the equation of motion is what is called the \textit{collision} term. Typical examples found in the literature following this strategy are Green Function \cite{Klimeck,Klimeck2,Green}, the density matrix \cite{Rossi1,Rossi2}, Wigner distribution function \cite{Ferry,Frensley,Dollfus1,Dollfus2,Nedjalkov,Jonasson,Polkovnikov}, the master equation \cite{Fischetti1,Fischetti2}, Kubo formalism \cite{Cummings}, conditional wave functions \cite{Bohm_original,Bohm1,Proceddings,PRLxavier,Entropy}, etc. 

There are two main difficulties in these quantum transport approaches with dissipation. The first difficulty is that open quantum systems cannot be described by an (orthodox) pure state, but by a density matrix (or some transformation of it). Thus, we cannot assign a pure state to an electron and we have difficulties to identify which are the properties of each electron that are modified by \textit{collisions}. The second difficulty appears because the collisions are, at best, a reasonable approximation of the real interaction between the simulated and non-simulated degrees of freedom (but never an exact result). Both difficulties make the evaluation of the physical soundness of a collision model for quantum transport a difficult task. A useful criteria for evaluating collisions models is checking about its complete positivity. The complete positivity means that the (reduced) density matrix will not provide negative values of the probability presence along the device. The presence of such negative values is unphysical because the typical probabilistic interpretation is precluded then \cite{open,vega}. The Wigner distribution function, as a Wigner-Weyl transformation of the density matrix \cite{Ferry,Wigner}, does also suffer from such lack of complete positivity. We clarify that we are not referring here to the fact that the Wigner function is a quasi-probability \cite{Zhen_true}, but to the fact of obtaining negative probability of finding electrons at some positions of the device. For example, it is well-known that the typical use of a Boltzmann-like superoperator in the description of the collision term in the Wigner formalism can produce such regions of negative probability presence \cite{Rossi1,Rossi2,Zhen}. Other issues like the lack of energy conservation have been underlined in the use of a Boltzmann superoperator, \cite{Rossi3,Nedjalkov2}.

In this work, we will focus on an additional criteria to verify the physical soundness of collision models. It is very reasonable to assume that most collisions between particles induce a well-defined change of the energy of the electron. The physical justification of such assumption is that an hypothetical exact solution of the interaction of the particles involved in a collision, through a many-body Schr\"odinger equation, would show that the total energy (the ensemble value of the Hamiltonian) is a constant of motion. However, in the density matrix (or some transformation of it), it is not always possible to access to the energy of each particle. Without such information, it is not clear if a collision model satisfies or not the mentioned energy conservation requirements. A clear example of such difficulties appears in the formulation of quantum transport through the Wigner distribution function. In the Wigner function, the information of the quantum system is given by one position and one momentum degree of freedom. No information of the energy of each individual particle is given by the Wigner formalism. One can assume that the kinetic energy of particles can be obtained from the value of their momentum. However, in many practical scenarios, there is no one-to-one relation between energy and momentum (the energy eigenstates are not momentum eigenstates, and viceversa). Then, a reasonable change in the momentum in the Wigner function after a collision even can be translated into an unexpected/uncontrolled change in the total energy, making the final result of the collision process unphysical.

After this introduction, in Sec. \ref{s2:0}, we will define the two conditions, complete positivity and energy conservation, for modeling collisions in a general density matrix formulation of a quantum system. In Sec. \ref{s3}, we will show an exact solution for the electron-photon interaction  emphasizing the importance of energy conservation during the electron-photon interaction. Then,  we will perform a comparison between approximate collision models that modify the energy of the electron and approximate collision models that modify the momentum of the electron, respectively. In Sec. \ref{s4}, we will show the Wigner-Weyl representation of such energy and momentum collisions models for several cases showing the non-physical results originated by the collisions models that exchange momentum in scenarios where energy and momentum do not commute. In Sec. \ref{s5}, we define the two practical requirements that are mandatory to be satisfied by a collision model implemented in the phase-space (Wigner) description to satisfy  complete positivity and energy conservation. In Sec. \ref{s5} we also add a subsection about time reversibility and how this influenced by collisions. We conclude in Sec. \ref{s6} indicating that alternative collision models with an explicit knowledge of the state ascribed to each electron seems strongly recommendable to avoid the previous two problems.

%%%%%%%%%%%%%%%%%%%%%%%%%%%%%%
%%%%%%%%%%%%%%%%%%%%%%%%%%%%%
\section{Problems in modelling collisions}
\label{s2:0}
%%%%%%%%%%%%%%%%%%%%%%%%%%%%%%
%%%%%%%%%%%%%%%%%%%%%%%%%%%%%%

The density matrix deals with mixed states which arise in quantum mechanics when the preparation of the system is not fully known, or when one wants to describe an open system which is entangled to an environment \cite{open}. Since both conditions are typically observed in electron devices, the density operator seems an adequate tool to study quantum transport. The typical equation of motion for the density operator $\hat \rho(t)$ is given by the Liouville-Von Neumann  equation
\begin{equation}
	\frac{d \hat\rho(t)}{d t}=\frac{1}{i \hbar}[\hat H, \hat \rho(t)]=\frac{1}{i \hbar}[\hat H_0, \hat \rho(t)]+\hat C[\hat \rho(t)].
	\label{Liouville2}
\end{equation}
The first term in the right hand side of \eqref{Liouville2} provide the unitary evolution of the systems with $H_0$ a single-particle Hamiltonian, while the collisions are introduced by the \textit{new} term given by the superoperator $\hat C$ acting on the density matrix operator. 

Due to a collision at time $t_s$, the new density matrix $\hat \rho^s(t_s+\Delta t)$ from \eqref{Liouville2} will be equal to the unitary density matrix with a free evolution until time $t_s$ given by $\hat\rho_0(t_s)=\hat\rho(t_s)+\frac{\Delta t}{i \hbar}[\hat H_0, \hat \rho(t_s)]$, plus some perturbation due to the collisions given by $\Delta \hat \rho(t_s)=\Delta t \hat C[\hat \rho(t_s)]$: 
\begin{eqnarray}
	\hat\rho^s(t_s+\Delta t)&=&\hat\rho(t_s)+\frac{\Delta t}{i \hbar}[\hat H_0, \hat \rho(t_s)]+\Delta t \hat C[\hat \rho(t_s)]\nonumber\\
	&=&\hat\rho_0(t_s)+\Delta \hat \rho(t_s).
	\label{Liouville3}
\end{eqnarray}
Hereafter, when needed, we will use the superindex $s$ to indicate elements that refer to the system after the scattering event has taken place. Under the assumption that changes in the state of the system need some time to occur (an example of such time delay will be provided in the exact solution of the electron-photon in Sec. \ref{s31}), a reasonable condition for the application of the model in \eqref{Liouville3} is a finite value of the scattering time $\Delta t$ \cite{Rossi3,Nedjalkov2,Gurov}.

In most computational algorithms, it is not possible to know which are the individual (pure) states that build the density matrix. However, such (microscopic) knowledge can be obtained within the computational technique known as stochastic Schr\"odinger equation (for Markovian systems) or within the Bohmian theory (for either Markovian or non-Markovian systems) \cite{deva,wiseman1,wiseman2}. Within such techniques, we can assume that the exact single-particle wave function $\psi_j(x,t)=\langle x|\psi_j(t)\rangle$ is known for each $j$-th electron to build the density operator $\hat \rho(t)$ as:
\begin{equation}
	\hat{\rho}(t)=\frac{1}{M} \sum_{j=1}^N M_j |\psi_j(t)\rangle  \langle \psi_j(t)|,
	\label{red_density_matrix}
\end{equation}
where $M_j$ is the number of states $|\psi_j(t) \rangle$ that are present in the system with $M=\sum_{j=1}^N M_j$ and $N$ the maximum number of possible types of states (to be able to deal with annihilation and creation of electronic states, $M_j$ can be incremented or decremented by one at the scattering time, $t_s$, but we do not write its time dependence explicitly to simplify the notation).   

To better understand the problems of complete positivity and energy conservation, in the next two subsections, we will discuss the change of the density matrix due to a collision in \eqref{Liouville3} by distinguishing between the algorithms that have access to the additional knowledge given by \eqref{red_density_matrix} and the ones that do not have access. The computation of collisions without such knowledge corresponds to most of the algorithms presented in the literature dealing with quantum transport. We will see that without the additional knowledge of \eqref{red_density_matrix} the collisions process implemented in \eqref{Liouville3} can lead to violations of the complete positivity and energy conservation. Such violations can easily be avoided using computational algorithms that have access to the additional information provided by \eqref{red_density_matrix}.

%%%%%%%%%%%%%%%%%%%%%%%%%%%%%%
%%%%%%%%%%%%%%%%%%%%%%%%%%%%%%

\subsection{The problem of complete positivity}
\label{s2:1}
%%%%%%%%%%%%%%%%%%%%%%%%%%%%%%
%%%%%%%%%%%%%%%%%%%%%%%%%%%%%%

For algorithms \cite{deva,wiseman1,wiseman2} that have access to the additional information in \eqref{red_density_matrix}, the interaction of one electron with a photon at time $t_s$, can be understood as an electron changing from its initial state $|\psi_{2} (t_s)\rangle$ to its final state $|\psi^s_{2} (t_s)\rangle$. Thus, the new scattered density matrix after the scattering is: 
\begin{eqnarray}
	\hat{\rho}^s(t)=\hat{\rho}_0(t)
	-\frac{1}{M}|\psi_2 (t)\rangle \langle\psi_2(t)|+\frac{1}{M} |\psi^s_{2}(t) \rangle \langle\psi^s_{2}(t)|.
	\label{wig2}
\end{eqnarray}
The last two terms in the above expression correspond to annihilating the old (description of the) electron $|\psi_{2}(t) \rangle$ and creating a new (description of the) electron $|\psi^s_{2}(t) \rangle$. Such collision corresponds to the new collision term in \eqref{Liouville3} as:
\begin{equation}
	\Delta \hat \rho(t)=-\frac{1}{M}|\psi_2 (t)\rangle \langle\psi_2(t)|+\frac{1}{M} |\psi^s_{2}(t) \rangle \langle\psi^s_{2}(t)|.
	\label{new}
\end{equation}
The presence density  (also known as the charge density) of the density matrix in \eqref{wig2} at any time $t > t_s$ can be computed as:
\begin{eqnarray}
	\langle Q(x,t) \rangle&=&\langle x |\hat \rho(t)| x \rangle =\sum_{j=1}^N \frac{M_j}{M}|\psi_j(x,t)|^2\nonumber\\&-&\frac{1}{M}|\psi_2(x,t)|^2+\frac{1}{M}|\psi^s_{2}(x,t)|^2.
	\label{q1}
\end{eqnarray}
Notice that $|\psi_2 \rangle$ and $|\psi^s_2\rangle$ are totally different states because the collision can change momentum, energy, spatial distribution of the probability, etc. The problem with collisions for those algorithms that have no access to the information in \eqref{red_density_matrix} is that the implementation of $\Delta \hat \rho(t)$ can correspond to subtracting a state that is not present in the initial density matrix $\hat{\rho}(t_s)$. Then, the term -$\frac{1}{M}|\psi_2(x,t)|^2$ can provide negative presence density $\langle Q(x,t) \rangle<0$ in some spatial regions or times, which is clearly unphysical. Certainly, the computational algorithms with the additional knowledge of \eqref{red_density_matrix} could avoid such unphysical result by just not performing the scattering process from $|\psi_{2}(t) \rangle$ to $|\psi^s_{2}(t) \rangle$ in \eqref{Liouville3} because such algorithms knows that there is no such initial state  $|\psi_{2}(t) \rangle$. The problem is that most of the computational algorithms dealing with quantum transport do not have access to such microscopic information from \eqref{red_density_matrix} and they have to blindly apply a change on the density matrix given by $\Delta \hat \rho(t)$ without knowing which are the implications of such change in terms of states. In other words, the study of quantum transport in terms of the (macroscopic) matrix $\rho(x,x')=\langle x' |\hat \rho(t)| x \rangle$ has the great computational advantage of not needing the (microscopic) knowledge of the states building such matrix, but it has the drawback of losing control on the physical meaning of a change of density matrix by an amount equal to $\Delta \rho(x,x')=\langle x' |\Delta \hat \rho(t)| x \rangle$ after a collision. 

A more formal discussion of this problem, without the previous classification in terms of knowledge of information given by \eqref{red_density_matrix}, shows that non-Lindblad type of collisions can violate complete positivity \cite{open,vega}. It is also well-known that the Boltzmann-like collision operator applied to the Wigner distribution function can violate such complete positivity \cite{Rossi1,Rossi2,Zhen}.

%%%%%%%%%%%%%%%%%%%%%%%%%%%%%%%%%%%%%%%%%%%%%%%%%%%%%%%%%%%%
%%%%%%%%%%%%%%%%%%%%%%%%%%%%%%%%%%%%%%%%%%%%%%%%%%%%%%%%%%%%
\subsection{The problem of energy conservation}
\label{s2:2}
%%%%%%%%%%%%%%%%%%%%%%%%%%%%%%
%%%%%%%%%%%%%%%%%%%%%%%%%%%%%%

We discuss now an additional problem, not related with the probability presence, but with the conservation of energy. For those algorithms \cite{deva,wiseman1,wiseman2} that have access to the information in \eqref{red_density_matrix}, the electron-photon collision means that an electron with (ensemble) energy $\langle E_2(t_s)\rangle=\langle \psi_2 (t_s)|\hat H_0|\psi_2 (t_s)\rangle$ at time $t_s$ change its energy, due to interaction with a photon of energy $E_\gamma$, to the new value $\langle E^s_2(t_s)\rangle=\langle \psi_2^s (t_s)|\hat H_0|\psi_2^s (t_s)\rangle$. The requirement of conservation of energy with such algorithms that have access to the microscopic states, $\langle E^s_2(t_s)\rangle=\langle E_2(t_s)\rangle+E_\gamma$, is trivially satisfied. 

The energy of the electron system at any time $t>t_s$ can be computed as:
\begin{eqnarray}
	\langle E^s(t) \rangle\rangle&=&\text{Tr}\left(\hat \rho^s(t) H_0 \right) =\sum_{j=1}^N \frac{M_j}{M}\langle E_j(t)\rangle \nonumber\\&-&\frac{1}{M}\langle E_2(t)\rangle+\frac{1}{M}\langle E^s_2(t)\rangle,
	\label{q2}
\end{eqnarray}
where $\langle E_j(t) \rangle =\langle \psi_j (t)|\hat H_0|\psi_j (t)\rangle$. Again, for algorithms that are not allowed to deal with the microscopic information of the state of each electron, before an after the collision, the change of energy due to the collision done in \eqref{Liouville3} has to satisfy: 
\begin{eqnarray}
	E_\gamma&=&\langle E^s_2(t)\rangle-\langle E_2(t) \rangle=\text{Tr}\left(\Delta \hat \rho(t) H_0 \right),
	\label{q3}
\end{eqnarray}
with $\Delta \hat \rho(t)$ defined in \eqref{new}. The problem is that it is not trivial to ensure that expression \eqref{q3} is satisfied, only using the macroscopic information that we have $\rho(x,x')=\langle x' |\hat \rho(t)| x \rangle$, and without knowing the information of the states involved in such collision.

In summary, as we will see along the paper, the computational advantage of the density operator of encapsulating all the physical information into the density matrix $\rho(x,x')=\langle x' |\hat \rho(t)| x \rangle$, avoiding to treat the heavy microscopic description of the state of each electron, also presents some drawbacks when dealing with collisions. The representation of the density matrix in coordinate space can be changed into the so-called Wigner distribution function $f_W(x,k,t)$ through the Wigner-Weyl transformation \cite{Ferry,Wigner}:
\begin{equation}
	f_W(x,k,t)=\frac{1}{2 \pi}\int dx' \, \, \, \langle x+\frac{x'}{2}|\hat \rho(t) |x-\frac{x'}{2}) \rangle \,\, e^{-ikx'}.
	\label{Wig_transf}
\end{equation}
Again, in general, the Wigner distribution function has no information at all about the microscopic states. Then, the collision operator can provoke unphysical violations of the conservation of energy, as we will see in following sections.

\section{Exact and approximate models for matter-light interaction}
\label{s3}

%%%%%%%%%%%%%%%%%%%%%%%%%%%%%%%%%%%%%%%%%%%%%%%%%%%%%%%%%%%%
%%%%%%%%%%%%%%%%%%%%%%%%%%%%%%%%%%%%%%%%%%%%%%%%%%%%%%%%%%%%
%%%%%%%%%%%%%%%%%%%%%%%%%%%%%%%%%%%%%%%%%%%%%%%%%%%%%%%%%%%%

We develop here an example of the role of the energy conservation in the collision of an electron with a photon. First, we explain and exact electron-photon Schr\"odinger equation and, then, we present two approximate collision models. The first approximate model for collisions will be based on changing the electron energy during the scattering process, while the second one is based on change of the value of momentum.

\subsection{Exact electron-photon interaction}
\label{s31}

The electromagnetic field in our quantum description is assumed to be a monochromatic field inside a optical cavity of length $L_{\gamma}$ with the shape $E(x,t) \propto q\;cos (k_\gamma x-\omega t)$ with $\omega$ the angular frequency of the field and $k_\gamma$ the wave vector giving a speed of light $c=\omega/k_\gamma$. The dependence of $E(x,t)$ on $x$ can be removed by assuming $L_x<<L_{\gamma} =\frac{2 \pi}{\omega}$ where $L_x$ is the length of the active region of the device, where electrons are simulated. The quantization of the electromagnetic field appears because the amplitude $q$ is not a fixed value, but a variable degree of freedom. Then, the Hamiltonian of the electromagnetic field in the $q$-representation $H_\gamma$ can be written as \cite{light}:
\begin{eqnarray}
	\hat	H_{\gamma} =-\frac{\hbar^2}{2} \frac{\partial ^2 }{\partial q^2}+\frac{\omega^2}{2}q^2.
	\label{H_gamma_q}
\end{eqnarray}
Inside the optical cavity, the vacuum state related to the absence of photons is $|0\rangle$, corresponding to the ground state of an harmonic oscillator $\psi_0(q)=\langle q | 0 \rangle$ of \eqref{H_gamma_q}. Identically, the first state of the harmonic oscillator in \eqref{H_gamma_q} corresponds to the presence of one mode, or one photon, inside the optical cavity described by $\psi_1(q)=\langle q| 1 \rangle$.
For more details, see \cite{Entropy}.

The electron part of the system follows the well known electron Hamiltonian, in the $x$-representation: 
\begin{equation}
	H_0=-\frac{\hbar^2}{2m} \frac{\partial^2}{\partial x^2}+V(x),
	\label{He}
\end{equation}
where $V(x)$ includes both internal and external scalar potential and $m$ is the electron effective mass. The Hamiltonian $H_0$ is written assuming a 1D electron system.

Finally, the electron-photon wavefunction $\Psi(x,q,t)$ will be guided by the following Schr\"odinger-like  equation of motion:
\begin{eqnarray}
	\label{scho}
	i \hbar \frac{\partial \Psi(x,q,t)}{\partial t} =& -&\frac{\hbar^2}{2 m} \frac{\partial ^2 \Psi(x,q,t)}{\partial x^2}+V(x)\Psi(x,q,t) \nonumber\\
	&-&\frac{\hbar^2}{2} \frac{\partial ^2 \Psi(x,q,t)}{\partial q^2}+\frac{\omega^2}{2}q^2\Psi(x,q,t)\nonumber\\
	&+&\alpha' x q \Psi(x,q,t),
	\label{total}
\end{eqnarray}
where the last term in \eqref{total} is the interaction term between the optical and electron parts in the typical dipole approximation \cite{light}. The parameter $\alpha'$ is the interaction strength.

Since we are only interested in a dynamic process involving only two possible photons states, $\psi_0(q)$ for zero photons and $\psi_1(q)$ for one photon, the whole wave function  $\Psi(x,q,t)$ can be decomposed as:
\begin{eqnarray}
	\label{super}
	\Psi(x,q,t) = \psi_A(x,t)\psi_0(q)+\psi_B(x,t)\psi_1(q),
\end{eqnarray}
with
\begin{eqnarray}
	\label{def_electron_side}
	\psi_{A/B}(x,t)=\int \; \psi_{0/1}^*(q)\Psi(x,q,t) dq.	
\end{eqnarray}
The wave functions $\psi_A(x,t)$ and $\psi_B(x,t)$ describe how the total wave function is projected in $\psi_0(q)$ and $\psi_1(q)$ respectively. 

The equation of motion of $\psi_A(x,t)$ and $\psi_B(x,t)$ can be obtained  by introducing the definition \eqref{super} into \eqref{scho} and using the orthogonality of $\psi_0(q)$ and $\psi_1(q)$ as follows\\
\begin{eqnarray}
	\label{schoA}
	i \hbar \frac{\partial \psi_A(x,t)}{\partial t} &&=-\frac{\hbar^2}{2 m} \frac{\partial ^2 \psi_A(x,t)}{\partial x^2}+ \\ &&\left(V(x)+\frac{1}{2}\hbar \omega\right)  \psi_A(x,t)+ \alpha x \psi_B(x,t)\nonumber
\end{eqnarray}
and
\begin{eqnarray}
	\label{schoB}	
	i \hbar \frac{\partial \psi_B(x,t)}{\partial t} &&=-\frac{\hbar^2}{2 m} \frac{\partial ^2 \psi_B(x,t)}{\partial x^2}+ \\&&\left(V(x)+\frac{3}{2}\hbar \omega\right)  \psi_B(x,t) +\alpha x \psi_A(x,t),\nonumber
\end{eqnarray}
where we have defined $\alpha=\alpha' \int \psi_0(q) q \psi_1(q) dq$ and we have used $\int \psi_0(q) q \psi_0(q)\;dq=\int \psi_1(q) q \psi_1(q)\;dq=0$. 

In this way the closed electron-photon system is now described with an exact model acting only on two electron wavefunctions. The present model can be extended to an open system through the use of the so-called Bohmian conditional wavefunction for electron wavefunctions $\psi_A(x,t)$ and $\psi_B(x,t)$. For explanation about Bohmian mechanics, we recall \cite{Bohm_original,Bohm1}, and for a definition and properties of the Bohmian conditional wavefunction see \cite{Proceddings,PRLxavier,Entropy}.

\subsection{Modelling electron-photon collisions as energy exchange}
\label{s32}

We consider now an electron defined by a single-particle state $\psi(x,t)$. As indicated above, for an open system like an electron device, such state can be understood as a Bohmian conditional wavefunction  \cite{Proceddings,PRLxavier,Entropy} where only the degree of freedom of the electron is considered $\psi(x,t) \equiv \psi(x,q(t),t)$ while the degree of freedom of the photon $q$ is fixed to some particular (Bohmian) value $q(t)$. See \cite{Entropy} for more details. 

At time $t_s$ the electron undergoes a collision with a photon. It is worth mentioning that the use of a finite time for the implementation of the scattering event is key to maintain continuity of the (conditional) wave function in space and time, but to simplify the discussion we will assume an instantaneous scattering process here. Later, in the numerical results, we will explicitly consider the finite duration of the scattering process. 

The process of collision $\psi(x,t_s) \to \psi^s(x,t_s)$ can be understood as a transition between the initial and final states $\psi(x,t_s)$ and $\psi^s(x,t_s)$, respectively, which satisfy $\langle E^s(t_s) \rangle=\langle E(t_s) \rangle + E_\gamma$, with $E_\gamma$ the energy of a photon. Within the energy representation, the wave packet can be decomposed into a superposition of Hamiltonian eigenstates $\phi_E(x)$ of the electron with hamitonian $\hat H_0$ in \eqref{He} as:
\begin{equation}
	\psi(x,t_s)=\int dE \; a(E,t_s) \; \phi_E(x),
\end{equation}
with $a(E,t_s)=\int dx \;\psi(x,t_s) \; \phi_E^*(x) $. The central energy $\langle E^s(t_s) \rangle$ after the scattering has to increase (or decreases)  an amount $E_\gamma$, so the new state after the collision can be written as:
\begin{eqnarray}
	\psi^s(x,t_s) &=&\int dE \; a(E - E_{\gamma},t_s) \; \phi_E(x) \nonumber\\&=& \int dE \; a^s(E,t_s) \; \phi_E(x),
	\label{scatt_E}
\end{eqnarray}
where we have defined $a^s(E,t_s)=a(E -E_\gamma,t_s)$.
This transition corresponds to absorption of energy by the electron. Emission can be identically modelled by using $\langle E^s(t_s) \rangle=\langle E(t_s) \rangle - E_\gamma$. In few words, the collision model performs a change in the wave packet central energy of an amount $\pm E_\gamma$.

%%%%%%%%%%%%%%%%%%%%%%%%%%%%%%%%%%%%%%%%%%%%%%%%%%%%%%%%%%%%
%%%%%%%%%%%%%%%%%%%%%%%%%%%%%%%%%%%%%%%%%%%%%%%%%%%%%%%%%%%%
%%%%%%%%%%%%%%%%%%%%%%%%%%%%%%%%%%%%%%%%%%%%%%%%%%%%%%%%%%%%

\subsection{Modeling electron-photon collisions as momentum exchange}
\label{s33}

It is straightforward to see that the process $\psi(x,t_s) \to \psi^s(x,t_s)$ satisfying a controlled change of the momentum can be done using $\psi^s(x,t_s)= e^{i p_\gamma x /\hbar} \psi(x,t_s)$, which  ensures that the state $\psi^s(x,t_s)$ has a increase of momentum with respect to $\psi(x,t_s)$ of $p_\gamma$. In fact, it can be easily demonstrated that such change on the momentum of $\psi^s(x,t)$ can be produced from the Schr\"odinger equation of the form:
\begin{eqnarray}
	i \hbar \frac{\partial \psi(x,t)}{\partial t} =& &\frac{1}{2m} \left(-i\hbar \frac{\partial}{\partial_x} +p_\gamma \Theta_{t_s} \right)^2 \psi(x,t) \nonumber\\
	&+&V(x)\psi(x,t),
	\label{schomomentum}
\end{eqnarray}
where $\Theta_{t_s}$ is a function equal to zero for $t<t_s$ and $1$  for $t>t_s$. Notice that the probability presence of the scattered wave packet satisfies $|\psi^s(x,t_s)|^2=|\psi(x,t_s)|^2$ because only a global phase $e^{i p_\gamma x /\hbar}$ is added. For a deeper explanation of the derivation of \eqref{schomomentum} we recall \cite{Enrique1}.

In an analogous way as the previous section, before the scattering, the state can be written as a superposition of momentum eigenstates $\phi_p(x)$ (which is also a basis of the electron in the x space) as:
\begin{equation}
	\psi(x,t)=\int dp \; b(p,t) \; \phi_p(x),
\end{equation}
with $b(p,t)=\int dx \; \psi(x,t) \; \phi_p^*(x) $. The central energy $\langle p(t) \rangle$ is increase of an amount $p_\gamma$, so in the momentum representation the state after the collision is:
\begin{eqnarray}
	\psi^s(x,t_s)&=&\int dp \; b(p-p_\gamma,t_s) \; \phi_p(x) \nonumber\\&=& \int dp \; b^s(p,t_s) \; \phi_p(x) \nonumber\\&=& e^{i p_\gamma x /\hbar} \psi(x,t_s),
	\label{scatt_p}
\end{eqnarray}
where we have defined $b^s(p,t)=b(p-p_\gamma,t)$. Notice that:
\begin{eqnarray}
	b(p-p_\gamma,t)&=&\int dx \; \psi(x,t) \; \phi_{p-p_{\gamma}}^*(x)\nonumber\\
	&=&e^{i p_\gamma x /\hbar} \int dx \; \psi(x,t) \; \phi_p^*(x).
\end{eqnarray}
Unfortunately, as it will be demonstrated in the following numerical results, a global mechanism of scattering valid for scenarios with potential barriers requires dealing with the energy conservation, and not with momentum conservation.

For the sake of clarity, the relationship of the approximated models in Sec. \ref{s32} and \ref{s33} with the exact model of Sec. \ref{s31} is here described. In the exact model, for each electron, both wavefunctions $\psi_A(x,t)$ and $\psi_B(x,t)$ are computed through \eqref{schoA} and \eqref{schoB}. This ensures the possibility of a continuous transition from $\psi_A(x,t)$ (zero photon) to $\psi_B(x,t)$ (one photon), and viceversa. In the approximate models, we just simulate one wave function $\psi(x,t)$ for each electron that takes into account the perturbation of the state due to scattering. In particular, in the case of a emission of a photon, the initial state is $\psi(x,0)=\psi_B(x,0)$ (one photon), and we wanted the final state at time $t_s$ provided by approximate model to be as similar as possible to $\psi^s(x,t_s) \approx \psi_A(x,t_s)$ (zero photon). For photon absorption, we have as the initial state $\psi(x,0)=\psi_A(x,0)$ (zero photon), and the final state $\psi^s(x,t_s) \approx \psi_B(x,t_s)$ (one photon).
\end{multicols}

\begin{figure*}[h]
	\hspace{-1cm}
		\begin{minipage}{\linewidth}
		{ \includegraphics[scale=0.22]{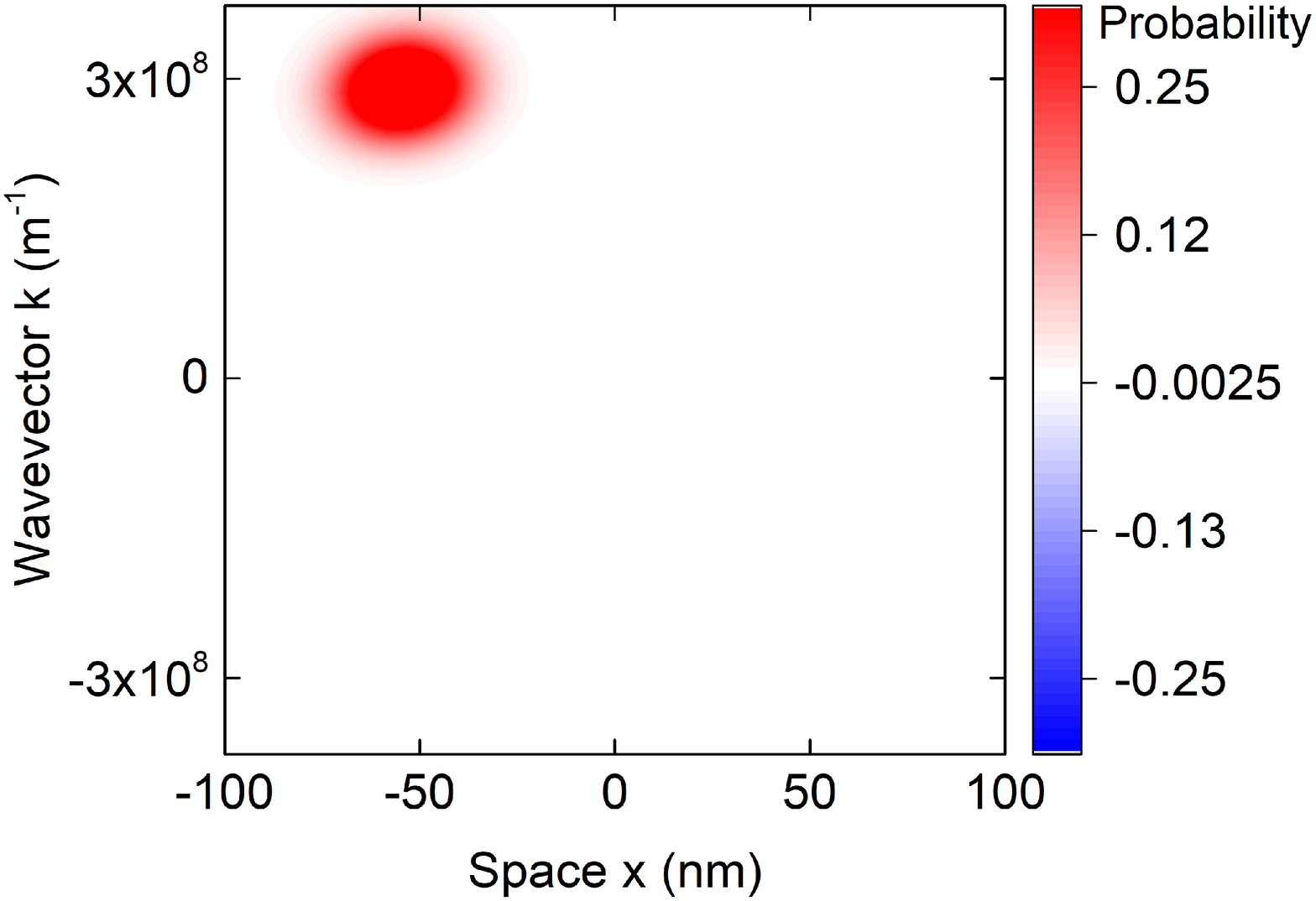}}
		{ \includegraphics[scale=0.22]{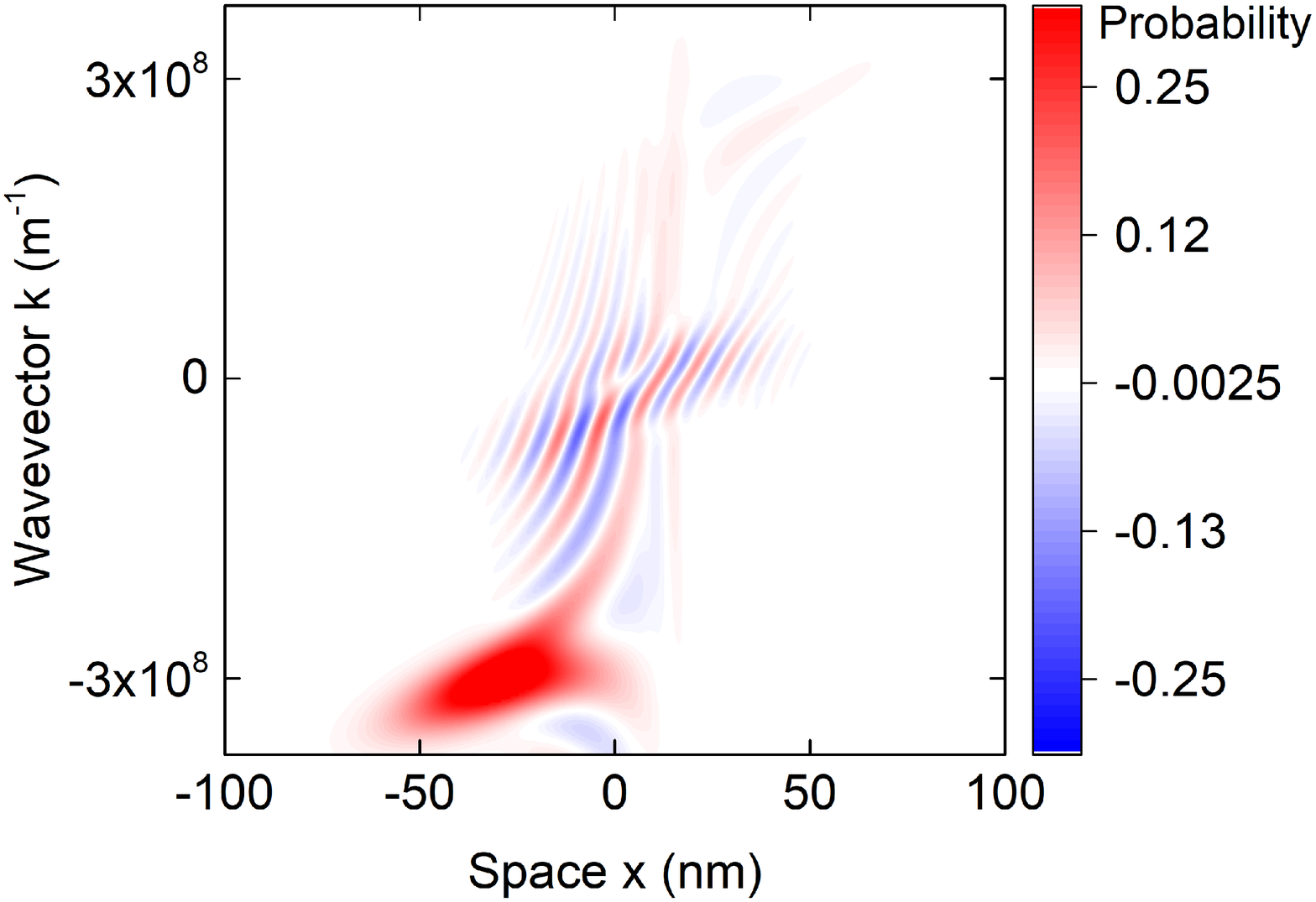}}
		{ \includegraphics[scale=0.22]{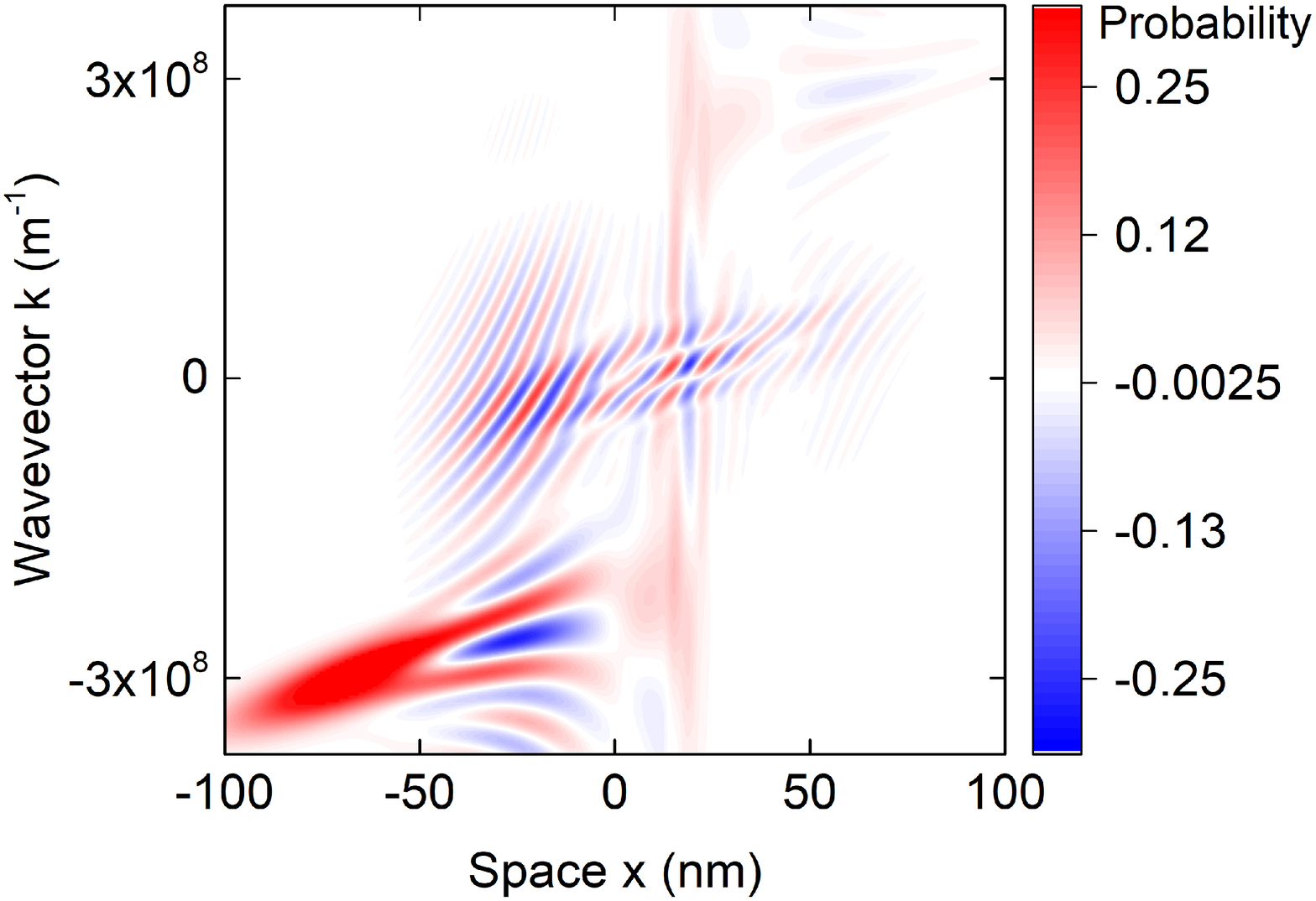}}\\
		{ \includegraphics[scale=0.22]{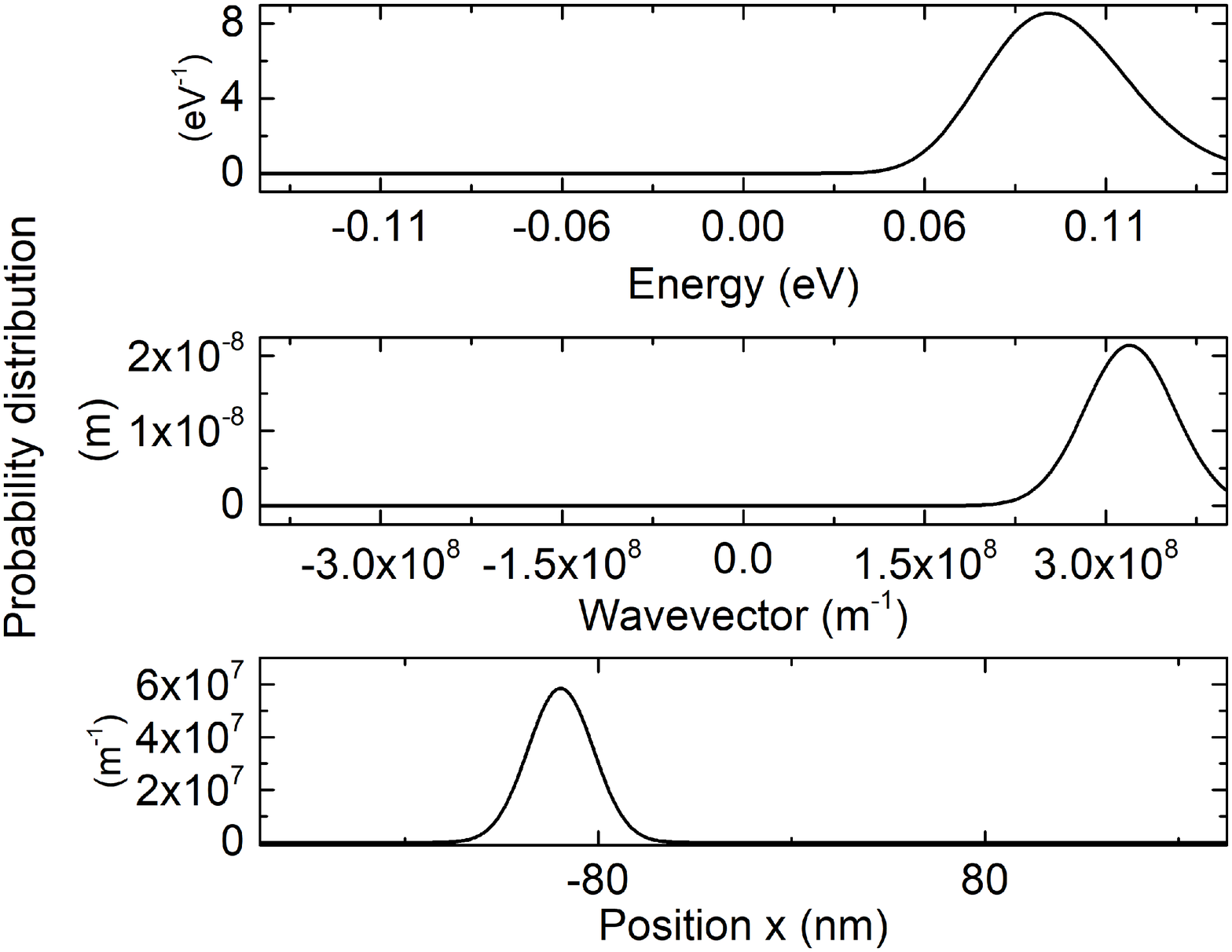}}
		{ \includegraphics[scale=0.22]{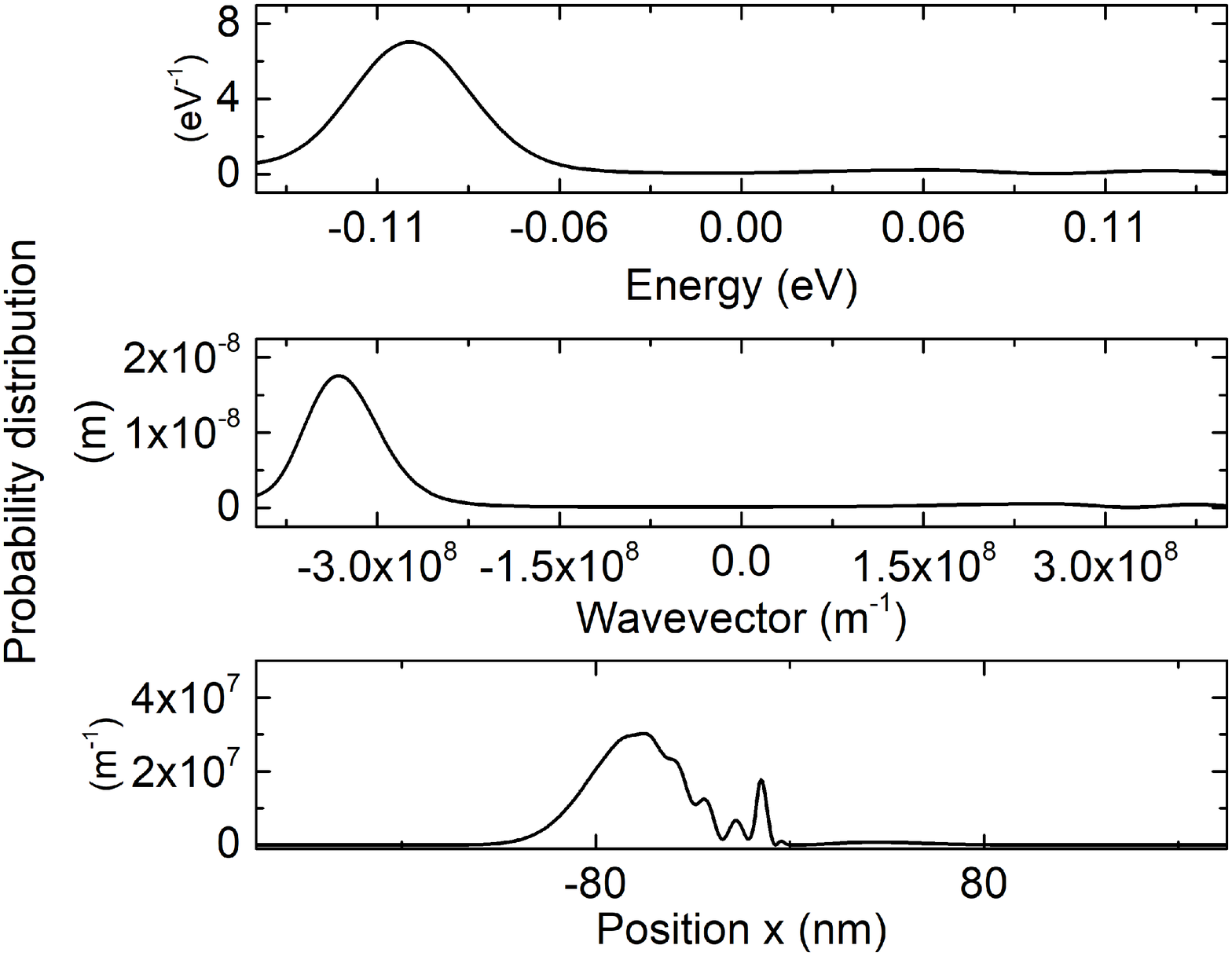}}
		{ \includegraphics[scale=0.22]{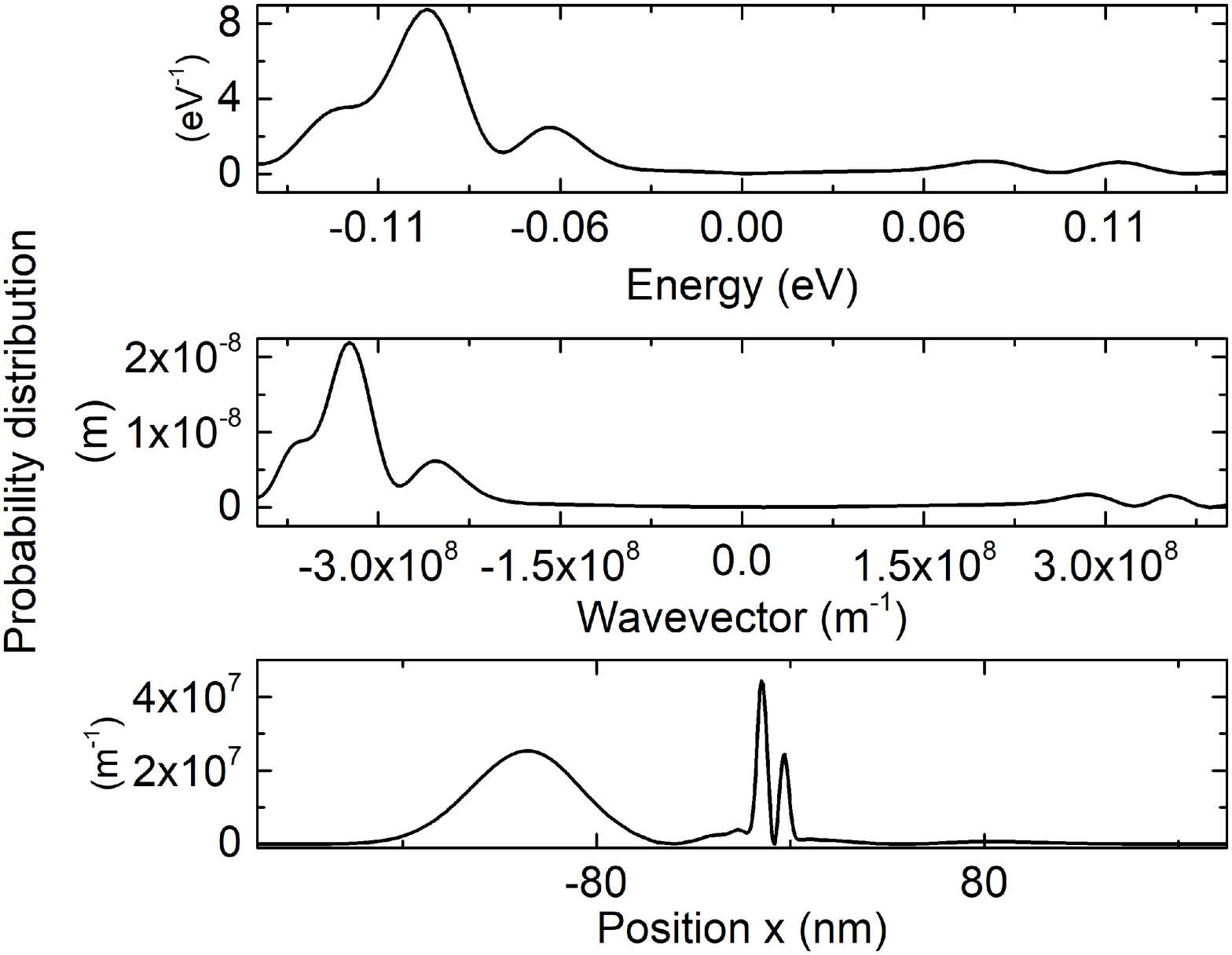}}
	\end{minipage}
	\caption{Wigner function of the total electron wavefunction interacting with a double barrier structure (green lines), at different times of its evolution: (a) at the beginning of the evolution, (b) when first entering inside the double barrier structure with $\langle E(t) \rangle=0.096 \, eV$, (c) after emitting a photon, so that the electron energy is $\langle E(t) \rangle=0.023 \, eV$. In (d), (e), (f) are shown projections along the energy (top), momentum (middle) and position (bottom) axis of the Wigner transform respectively of (a), (b), and (c).}
	\label{Exact_DB}
\end{figure*}

\begin{multicols}{2}

%%%%%%%%%%%%%%%%%%%%%%%%%%%%%%%
%%%%%%%%%%%%%%%%%%%%%%%%%%%%%%%
%%%%%%%%%%%%%%%%%%%%%%%%%%%%%%%
%%%%%%%%%%%%%%%%%%%%%%%%%%%%%%%
%%%%%%%%%%%%%%%%%%%%%%%%%%%%%%%

\section{Numerical results}
\label{s4}

In this section, we will compare the exact collision approach developed in the previous section, with the two collisions models mentioned there. Apart from plotting the evolution of the involved states before and after the collisions, we will also plot the Wigner distribution through the Wigner-Weyl transform defined from Eq. \eqref{Wig_transf}. In particular, we will consider scenarios where the electrons are suffering emission and absorption of photons, while impinging on a resonant tunneling diode. Such structure is composed by a double barrier with thickness 2nm, height 0.3 eV and a distance between the two barriers of 16 nm. The electron has effective mass $m=0.041 \, m_0$, where $m_0$ is the mass of the electron at rest. The resonant energies of the double barrier structure are $0.023$ eV for first resonant state and $0.096$ eV for the second resonant state.  

\subsection{Exact evolution of the Wigner Function}
\label{s40}
In this subsection we will show electron wave functions as defined in \eqref{super}. In particular, we start by considering $\psi_A(x,0)$ as a Gaussian wave packet in the left contact outside the barrier with an energy equal to the second resonant energy, while $\psi_B(x,0)=0$, meaning that initially there is no photon in the structure. Since both electron wavefunctions have a coupled evolution in \eqref{schoA} and \eqref{schoB}, step by step, the wave function $\psi_B(x,t)$ grows and $\psi_A(x,t)$ decreases indicating that a photon is being created. The Wigner function of the whole process is computed as the sum of the Wigner function linked to $\psi_A(x,t)$ (with \eqref{Wig_transf}) plus the Wigner function linked to $\psi_B(x,t)$ (again with \eqref{Wig_transf}). Notice that, from \eqref{super}, we get $\int dq |\Psi(x,q,t)|^2=|\psi_A(x,t)|^2+|\psi_B(x,t)|^2$ because $\psi_0(q)$ and $\psi_1(q)$ are orthogonal. The Wigner function is plotted in Fig. \ref{Exact_DB}, from (a) to (c) in three different times corresponding to the electron impinging upon a double barrier structure with an (initial) energy equal to the second resonant level of the double barrier. Inside the well the electron is interacting with the electromagnetic field and Rabi oscillations of the electron between first and second resonant levels are observed.
This can be seen from the wavefunction that, showing one maximum in the probability distribution inside the quantum well, is occupying the first level in 1(b) and (e). On the other hand, in Fig \ref{Exact_DB}(c) and (f) two maxima are observable inside the well (see (f)), so that the second level of the double barrier is occupied.

\subsection{Approximate evolution of the Wigner Function in free space}
\label{s41}
In this subsection we show the evolution of the Wigner function defined from a single electron evolving in free space and undergoing scattering through the (approximate) collision models explained in the Sec. \ref{s32} and \ref{s33}. The goal of this section is to prove the equivalence of two previous collisions models (energy exchange and momentum exchange) in the case of an electron system in free space. In free space, the momentum of a photon is much smaller than the momentum of an electron and the conservation of momentum (not only of energy) has to be satisfied. Thus, in this Sec. \ref{s41}, from a physical point of view, we consider the interaction of an electron with a phonon (instead of a photon).

\begin{figure*}[h]
	\hspace{-0.5cm}
			\begin{minipage}{\linewidth}
		{ \includegraphics[scale=0.35]{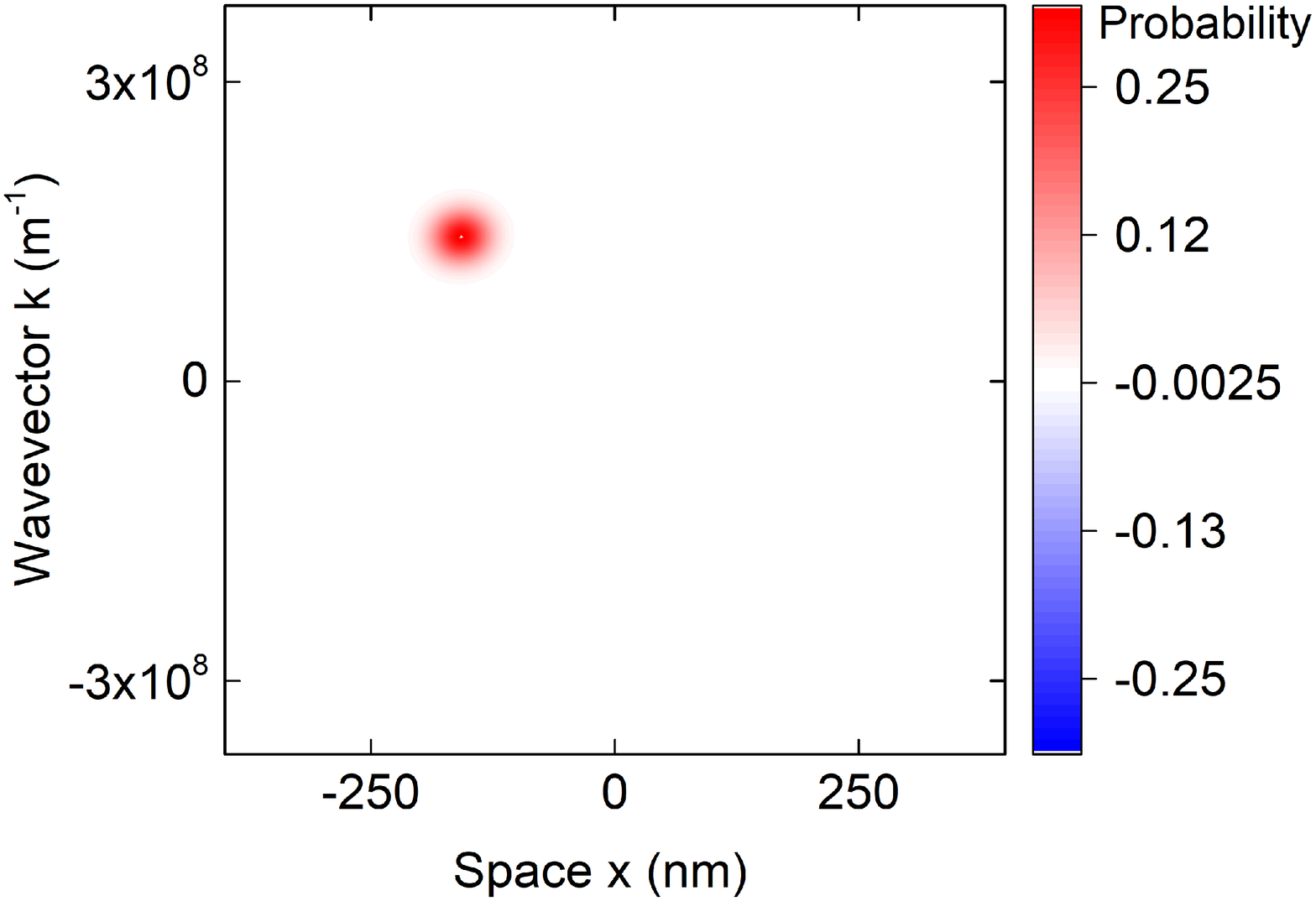}}
		{ \includegraphics[scale=0.35]{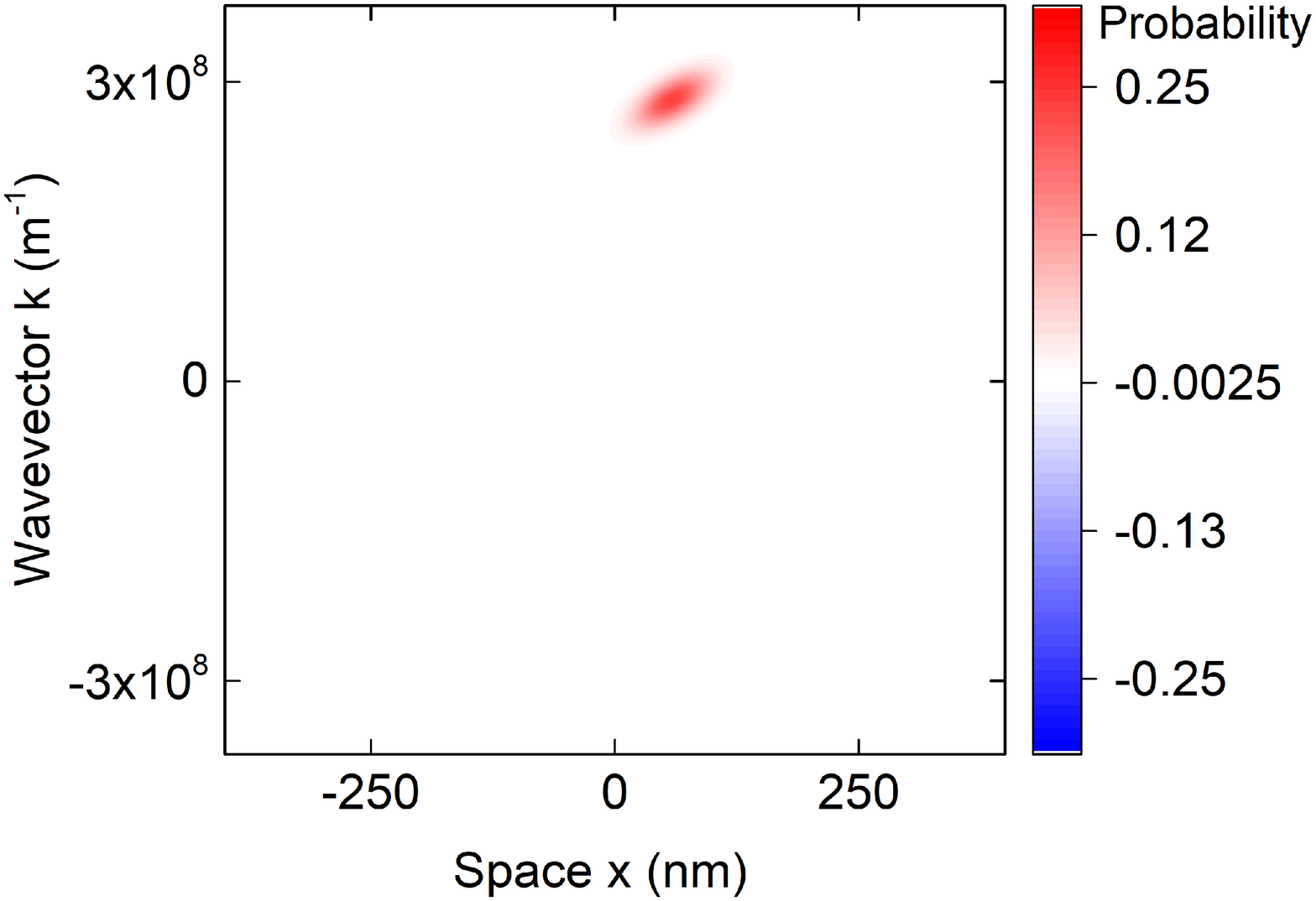}}\\
		{ \includegraphics[scale=0.35]{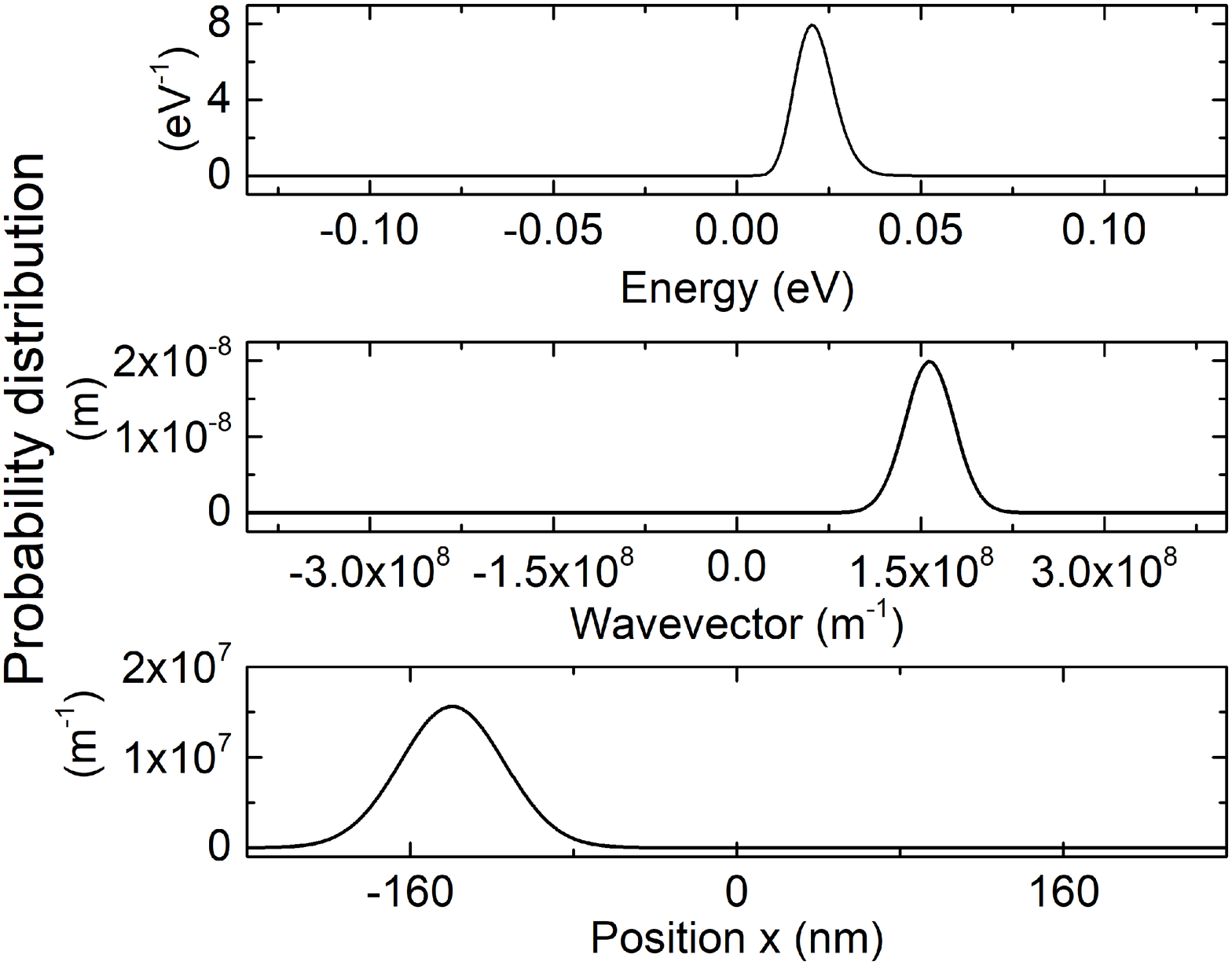}}
		{ \includegraphics[scale=0.35]{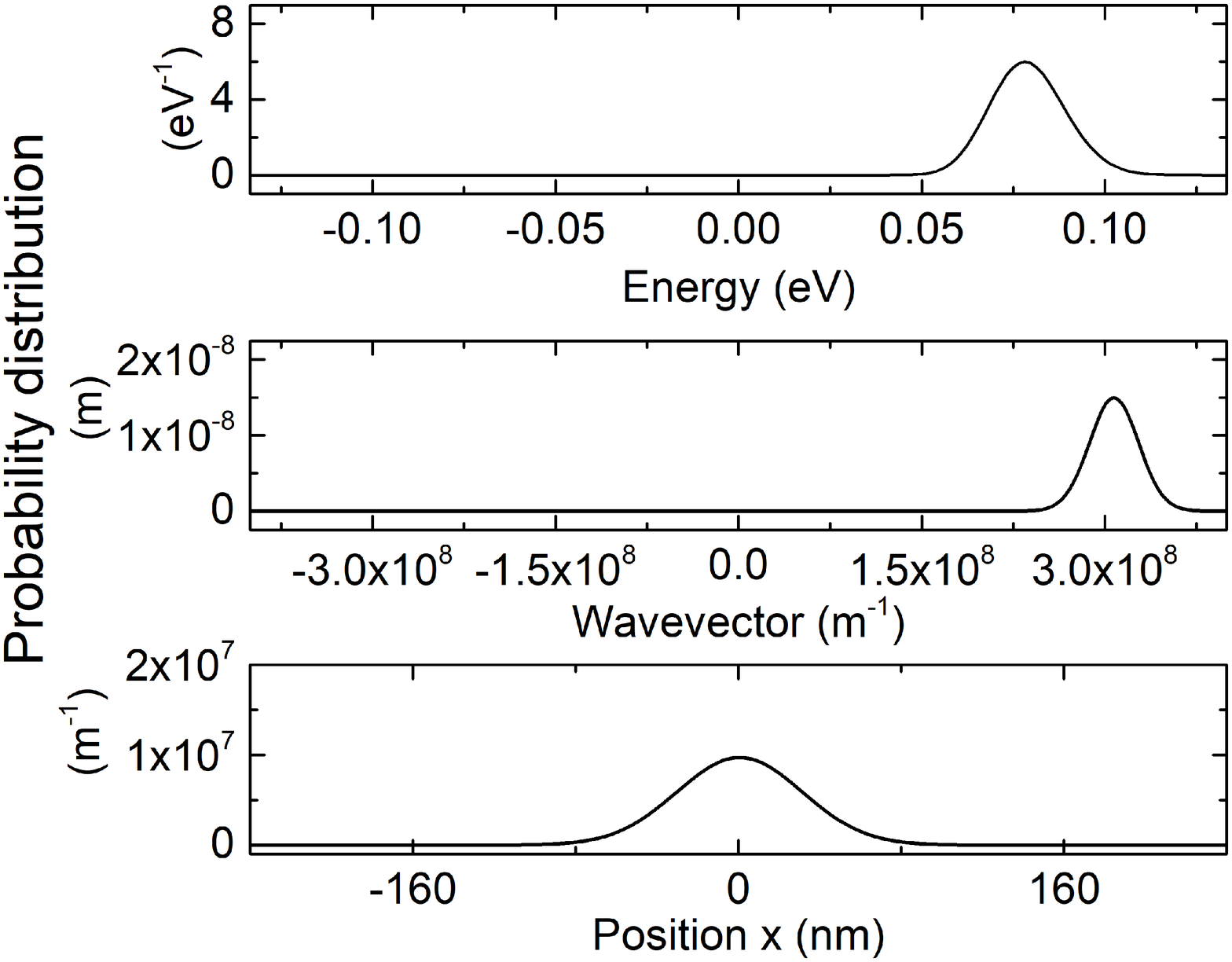}}
		\end{minipage}
	\caption{Wigner function for photon absorption in free space: (a) before scattering (b) scattered with the energy exchange and the momentum exchange model, which is equivalent to the former in free space conditions. In (c) and (d) are shown projections along the energy (top), momentum (middle) and position (bottom) axis of the Wigner functions respectively of (a) and (b).}
	\label{comparison_FP_pos}
	\end{figure*}

In \eqref{Liouville3}, is has been shown \cite{Rossi3,Nedjalkov2,Gurov} that the time of scattering $\Delta t$ has to be finite. In the present work, the transition shown in \eqref{scatt_E} for a scattering of energy $E_s$, is divided into $40$ steps of energy change $\Delta E_s=E_s/40$. Between two steps, the system is evolved for a  time of around $6$ fs. As a result, the total scattering time for the approximate model will be $40\cdot 6 = 240$ fs. An analogous implementation is done for \eqref{scatt_p}.
In Fig. \ref{comparison_FP_pos}, the electron is injected with an central energy equal to the first resonant level, $\langle E(t) \rangle=0.023$ eV, meaning a momentum of $\langle p(t) \rangle=1.573 \cdot 10^8 $ m$^{-1}$, This electron undergoes scattering for a finite time of approximately 0.1 ps to a final wave packet given by the collision models in Sec. \ref{s32} or Sec. \ref{s33} with an expectation value of the final energy $\langle E^s(t) \rangle=0.096$ eV, corresponding to a momentum of $\langle p^s(t) \rangle=3.214 \cdot 10^8  $ m$^{-1}$. This process $\psi(x,t_s) \to \psi^s(x,t_s)$ mimics, for example, an electron absorbing a phonon. The Wigner function is computed from \eqref {Wig_transf} for the states $\psi(x,t_s) \to \psi^s(x,t_s)$. 

\begin{figure*}[h]
		\hspace{-0.5cm}
		\begin{minipage}{\linewidth}
			{ \includegraphics[scale=0.35]{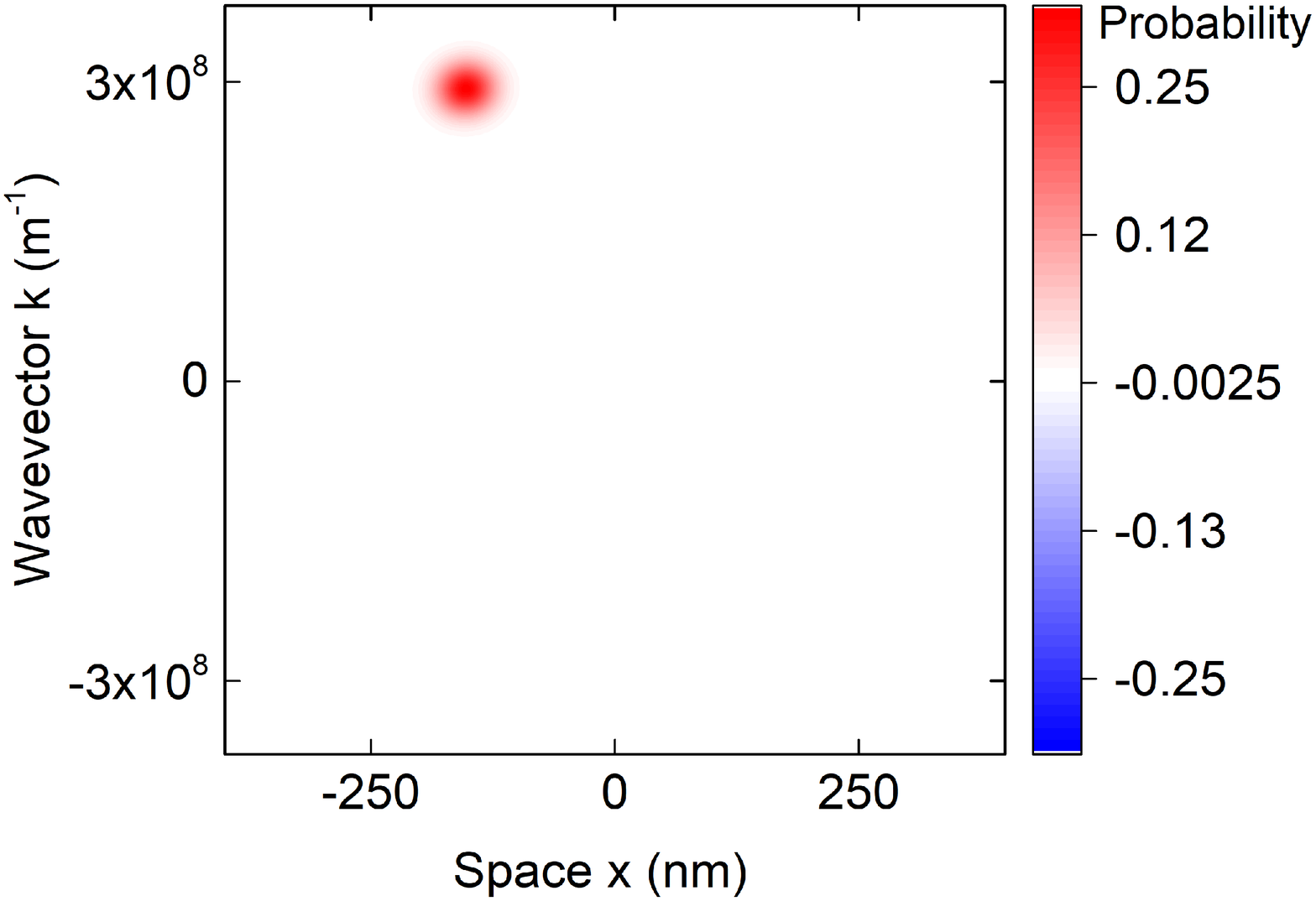}}
			{ \includegraphics[scale=0.35]{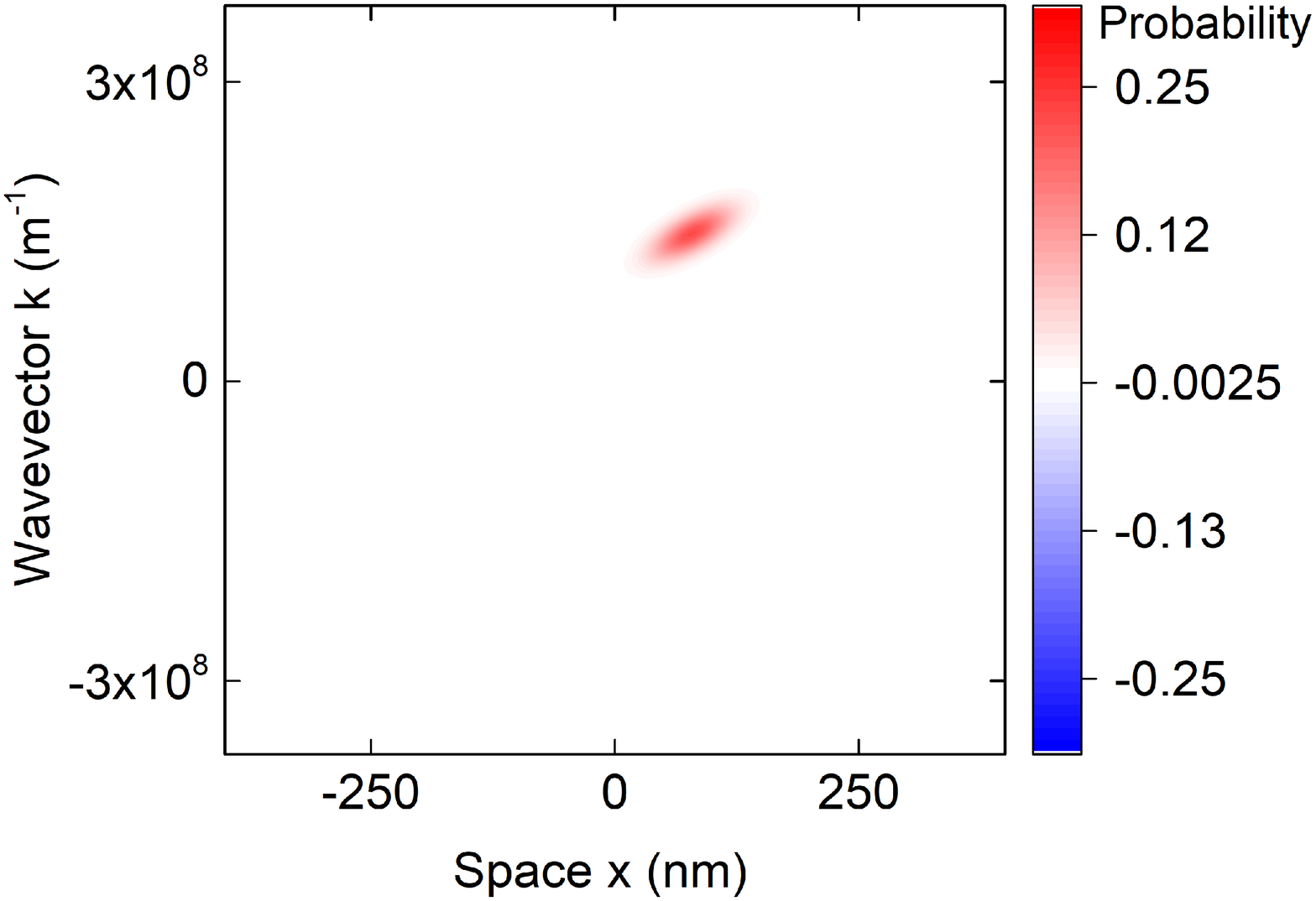}}
			{ \includegraphics[scale=0.35]{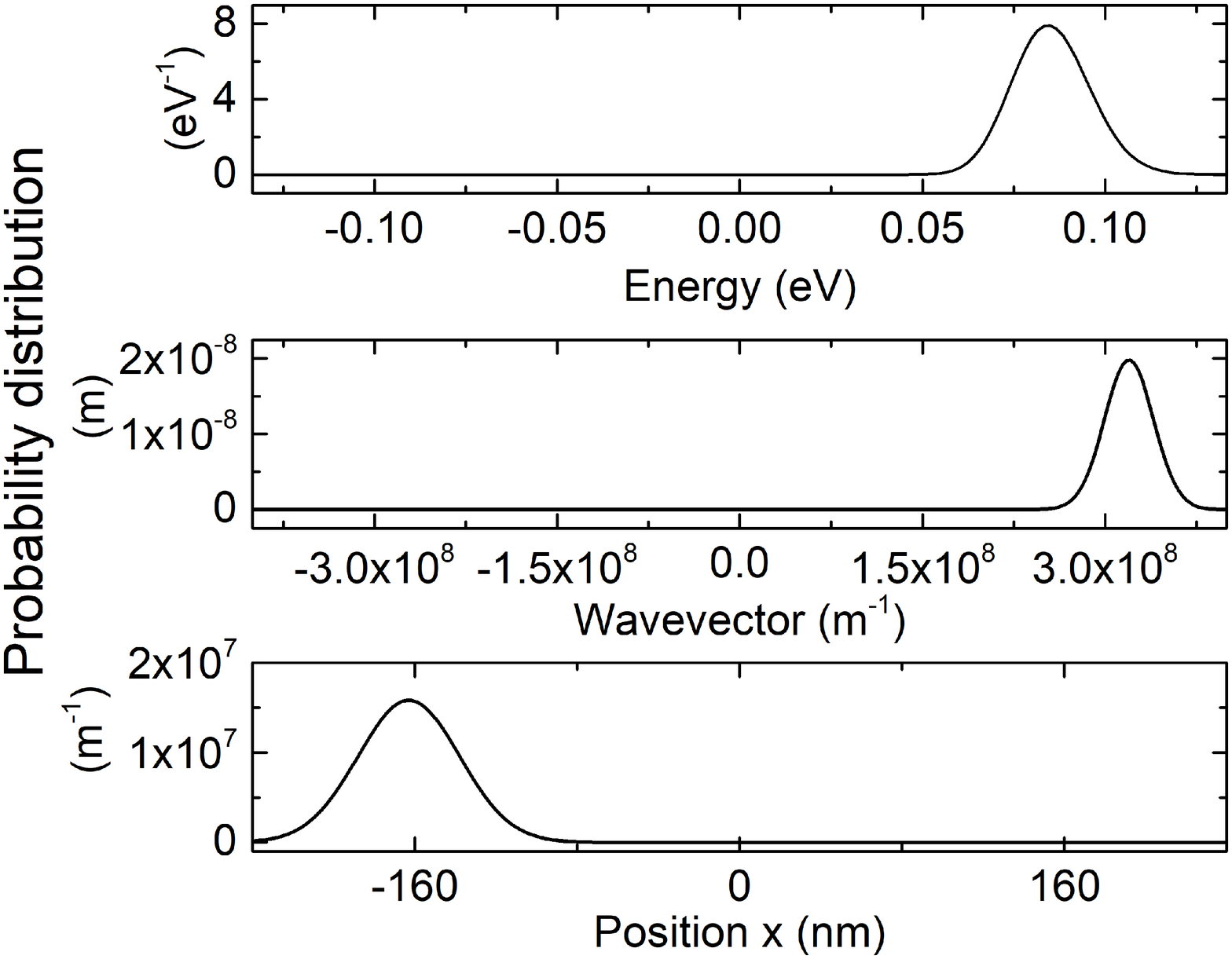}}
			{ \includegraphics[scale=0.35]{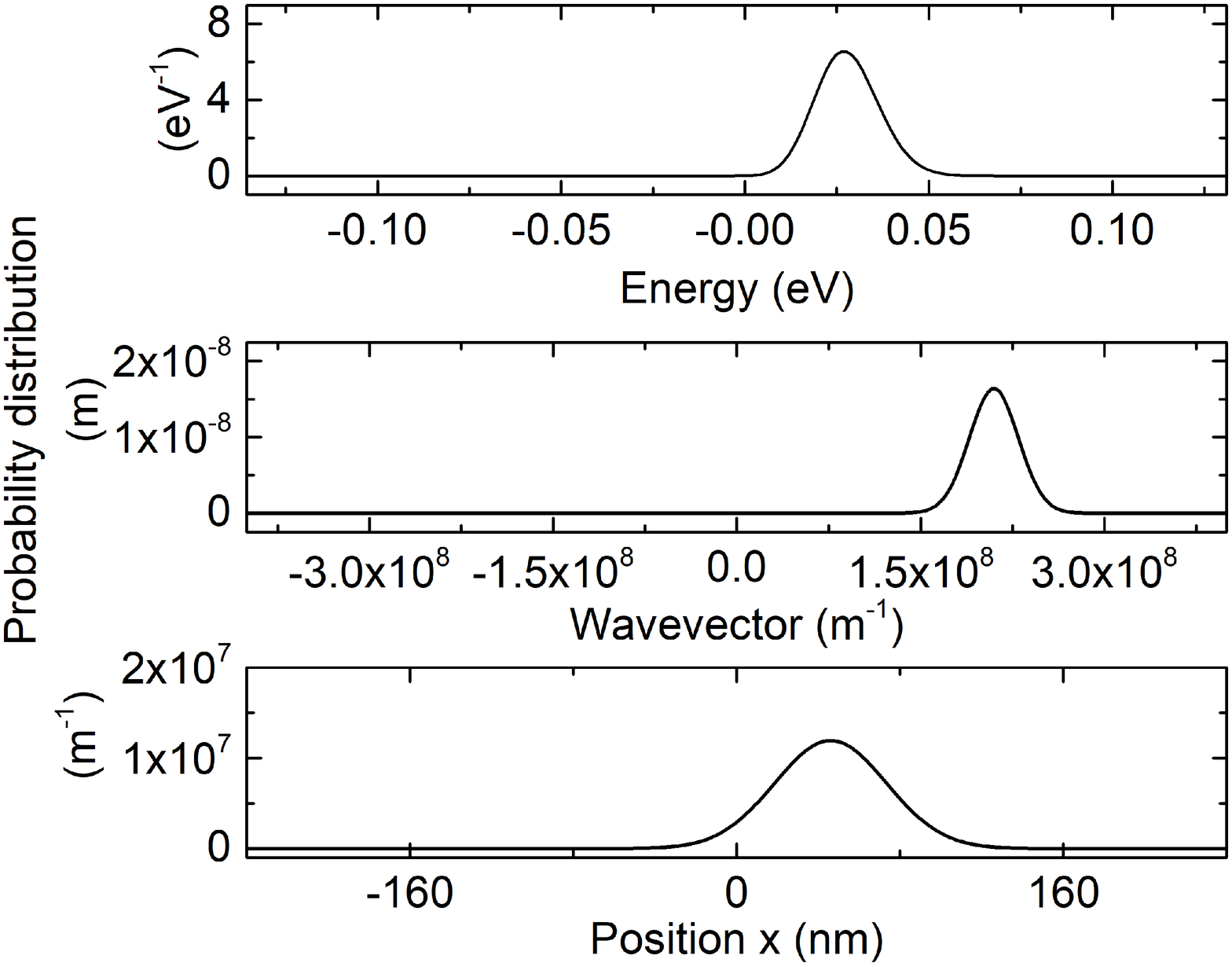}}
		\end{minipage}
		\caption{Wigner function for photon emission in free space: (a) before scattering (b) scattered with the energy exchange and the momentum exchange models, which is equivalent to the former in free space conditions. In (c) and (d) are shown projections along the energy (top), momentum (middle) and position (bottom) axis of the Wigner functions respectively of (a) and (b).}
		\label{comparison_FP_neg}
	\end{figure*}

In Fig. \ref{comparison_FP_neg}, the electron is injected with initial central energy $\langle E(t) \rangle=0.096$ eV, or momentum of $\langle p(t) \rangle=3.214 \cdot 10^8 $ m$^{-1}$. This electron also undergoes scattering in a finite time of approximately 0.1 ps and emits a phonon reaching a central energy of the wave packet equal to $\langle E^s(t) \rangle=0.023$ eV, corresponding to a momentum of $\langle p^s(t) \rangle=1.573 \cdot 10^8 $ m$^{-1}$. This process corresponds to an emission of phonon and the transition  $\psi(x,t_s) \to \psi^s(x,t_s)$ is computed from the collision models in Sec. \ref{s32} or Sec. \ref{s33} and the Wigner function linked to such states from \eqref {Wig_transf}. 

%\begin{figure}[h]

%\end{figure}
We conclude that the evolution of the Wigner function of an electrons scattered in free space behaves equivalently with both models, the one of Sec. \ref{s32} with energy conservation and the one of Sec. \ref{s33} with momentum conservation. The physical reason of such agreement is because, in free space, there is a one-to-one correspondence between the eigenvalues of momentum and energy operators (since these commute with each other). 
\end{multicols}
	
	\begin{multicols}{2}

\subsection{Approximate evolution of the Wigner Function with potential barriers}
\label{s42}

In this subsection, we show a comparison between the energy exchange and momentum exchange scattering models, respectively explained in Sec. \ref{s32} and \ref{s33}, for an electron wavefunction interacting with the double barrier structure described at the beginning of Sec. \ref{s4}. The momentum of an electron inside a quantum well tends to be very small (the wave function tends to become real) and the conservation of momentum is not a requirement because of the translation symmetry is broken by the barriers. Thus, in this Sec. \ref{s42}, we do consider the interaction of an electron with a photon. 

\begin{figure*}[h]
			\hspace{-0.5cm}
	\begin{minipage}{\linewidth}
		{ \includegraphics[scale=0.23]{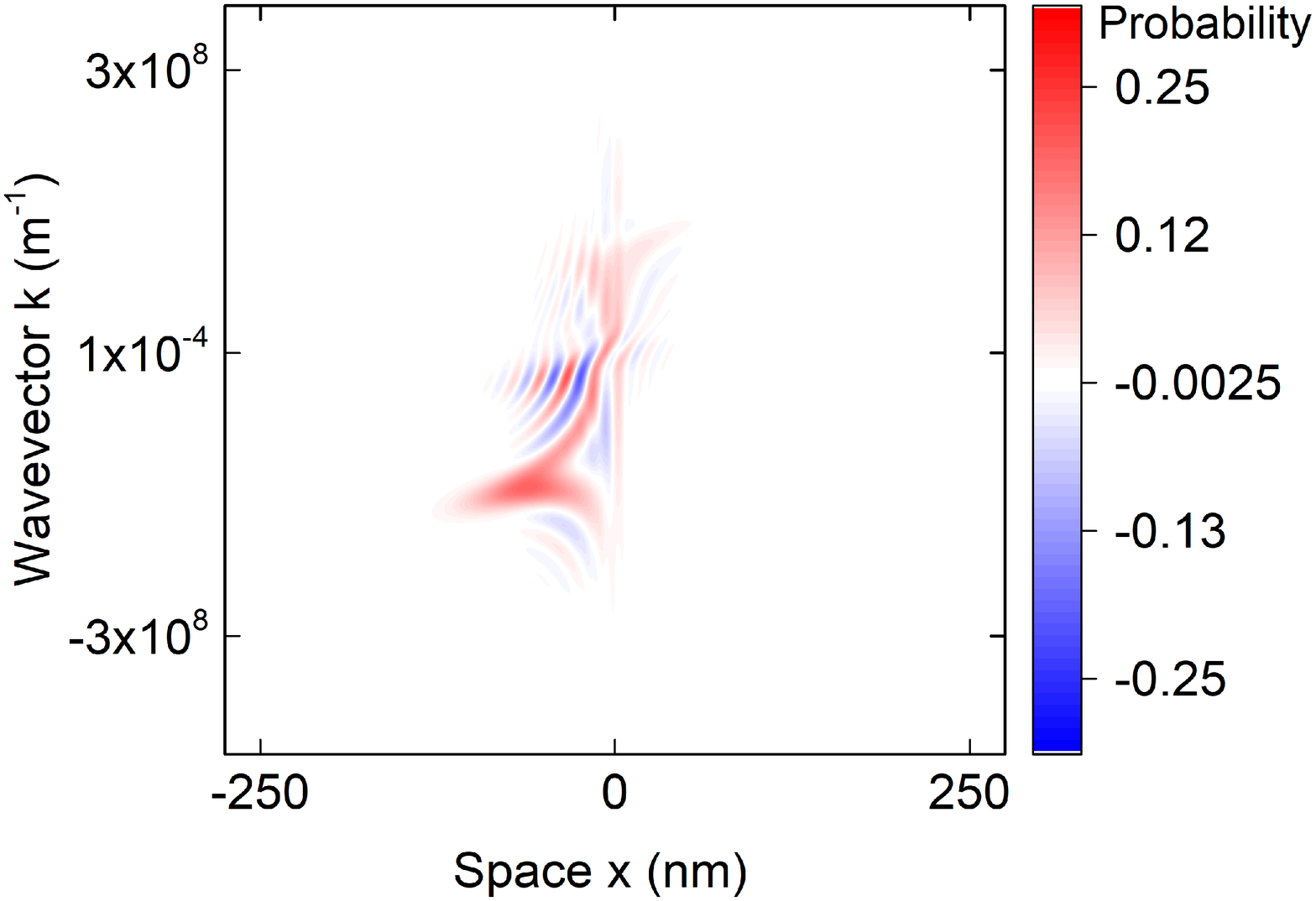}}
		{ \includegraphics[scale=0.23]{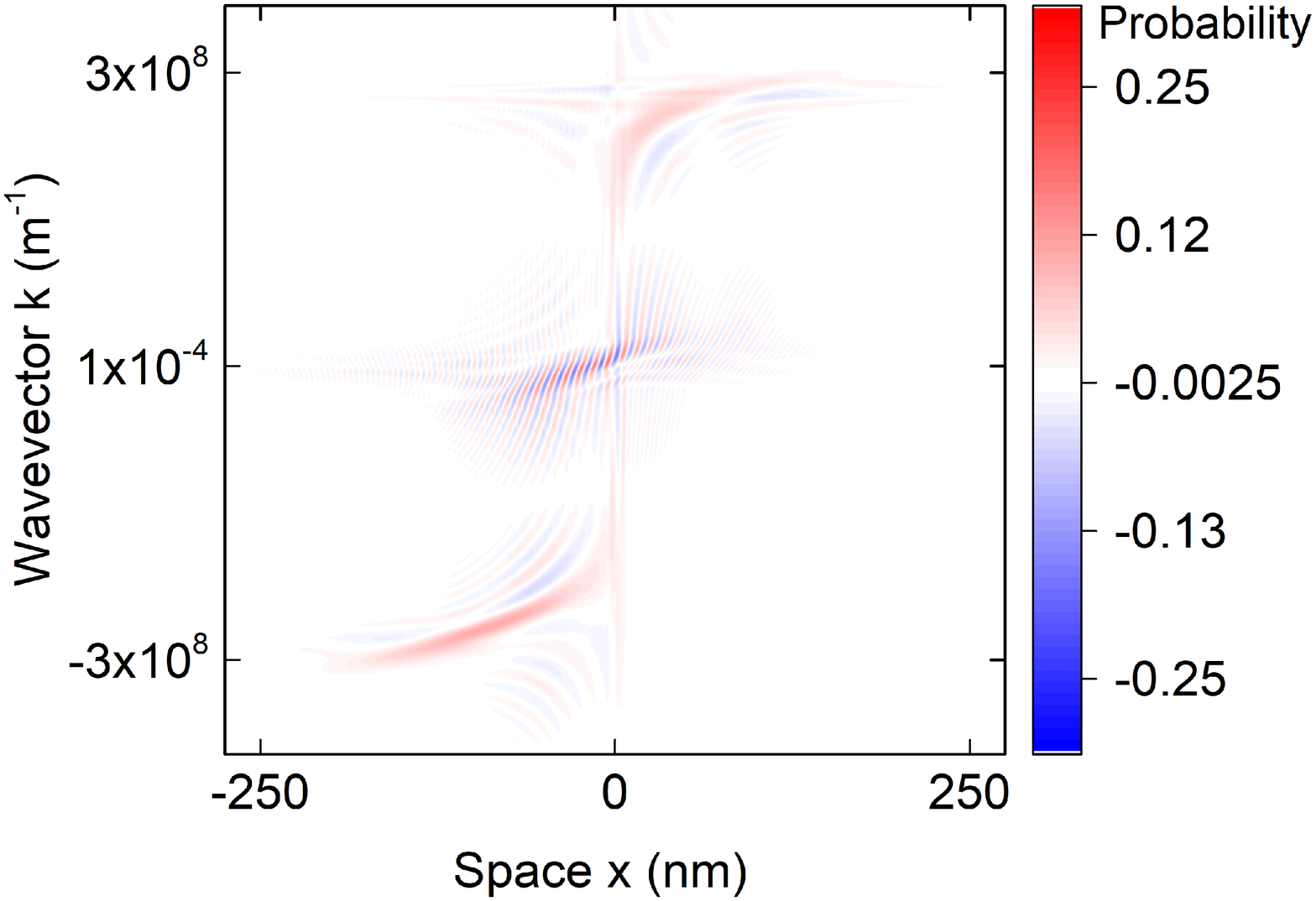}}
		{ \includegraphics[scale=0.23]{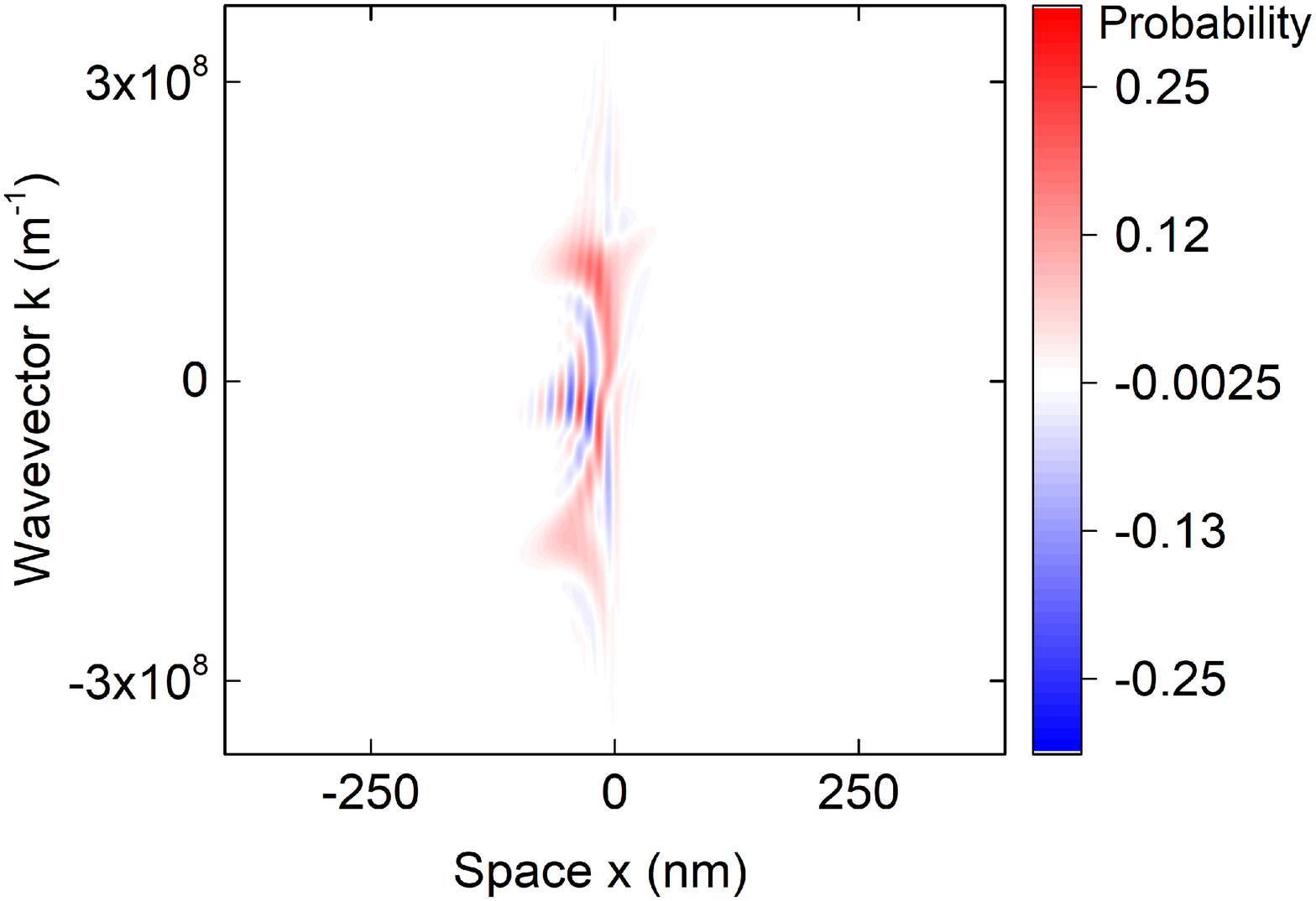}}
		{ \includegraphics[scale=0.23]{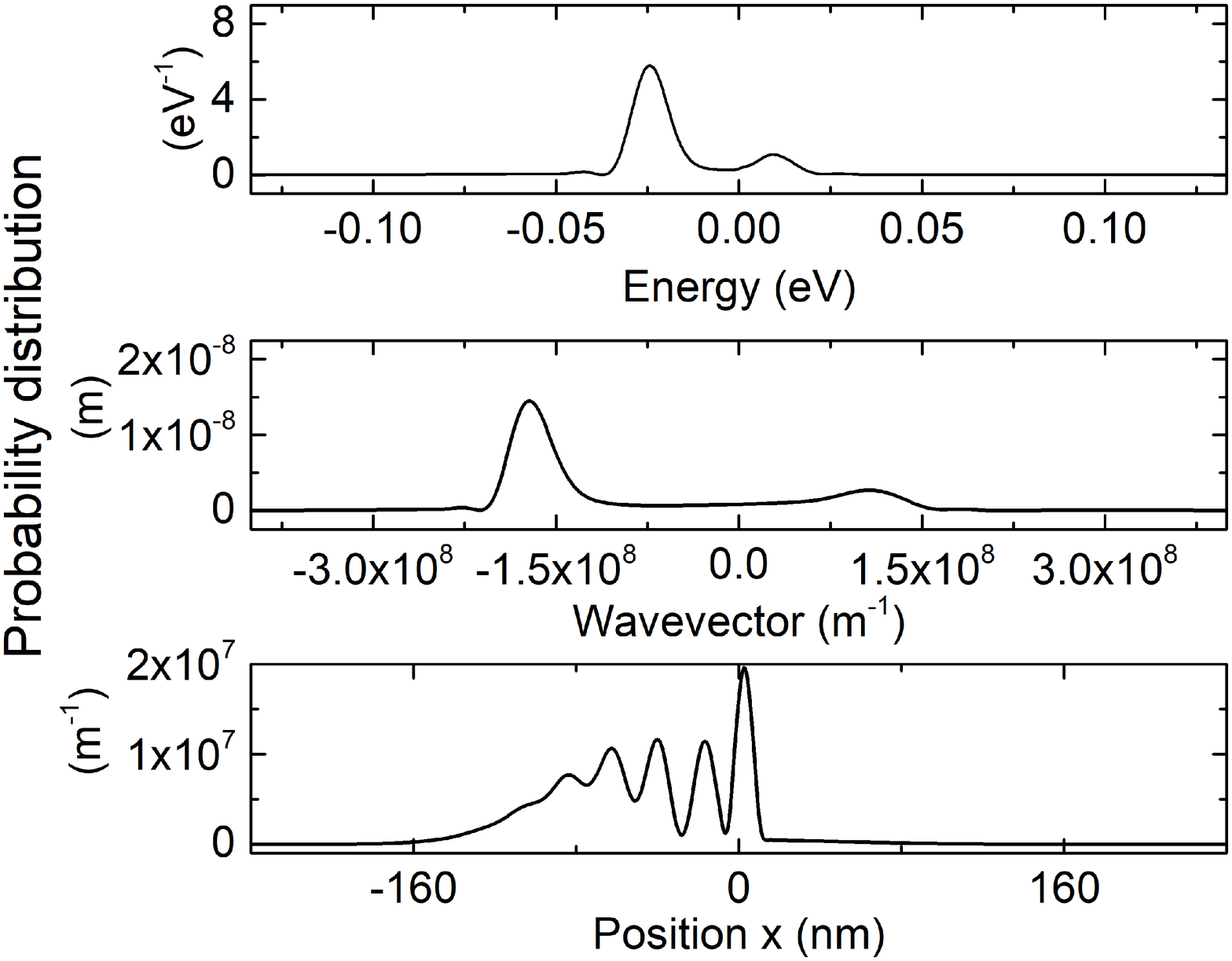}}
		{ \includegraphics[scale=0.23]{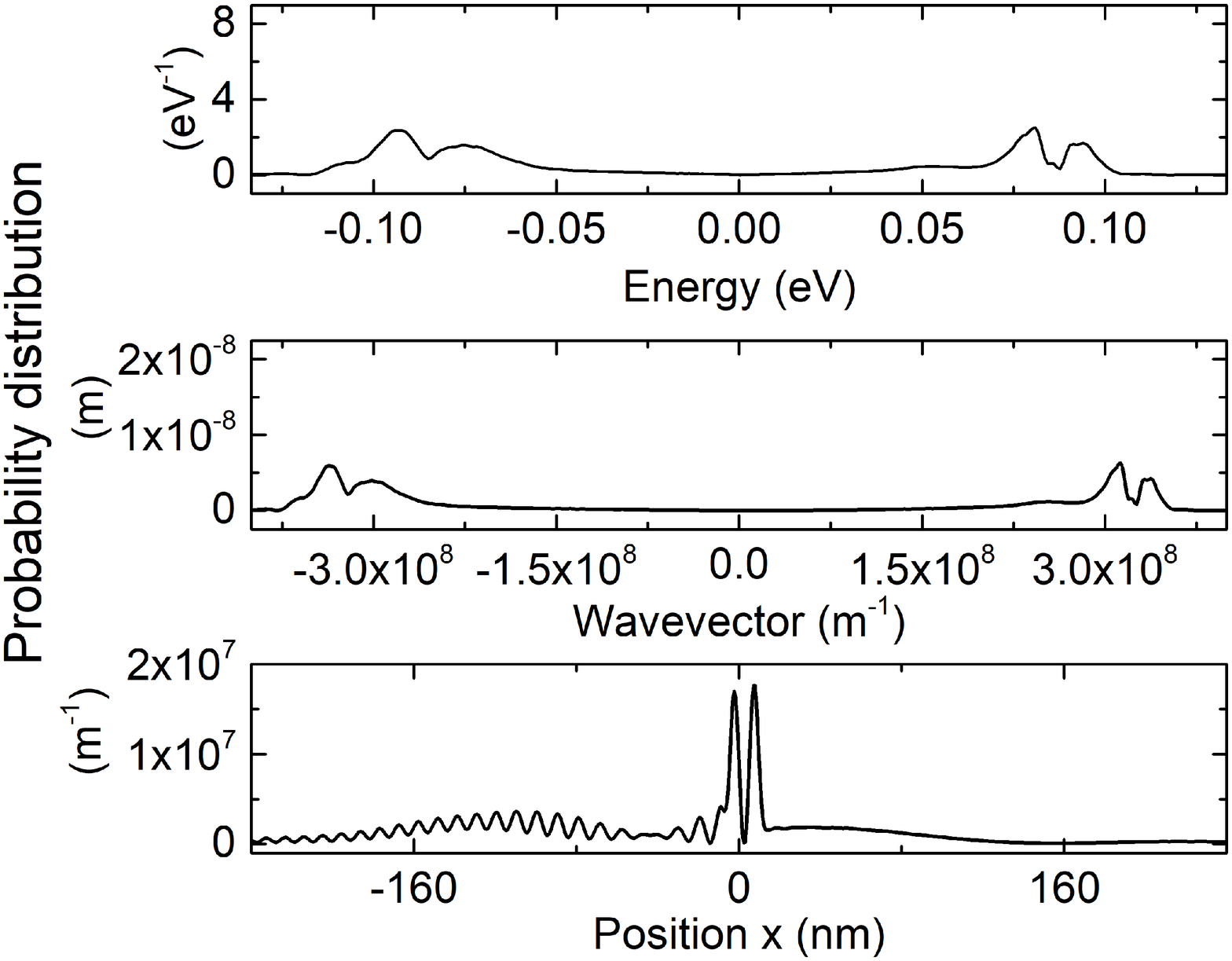}}
		{ \includegraphics[scale=0.23]{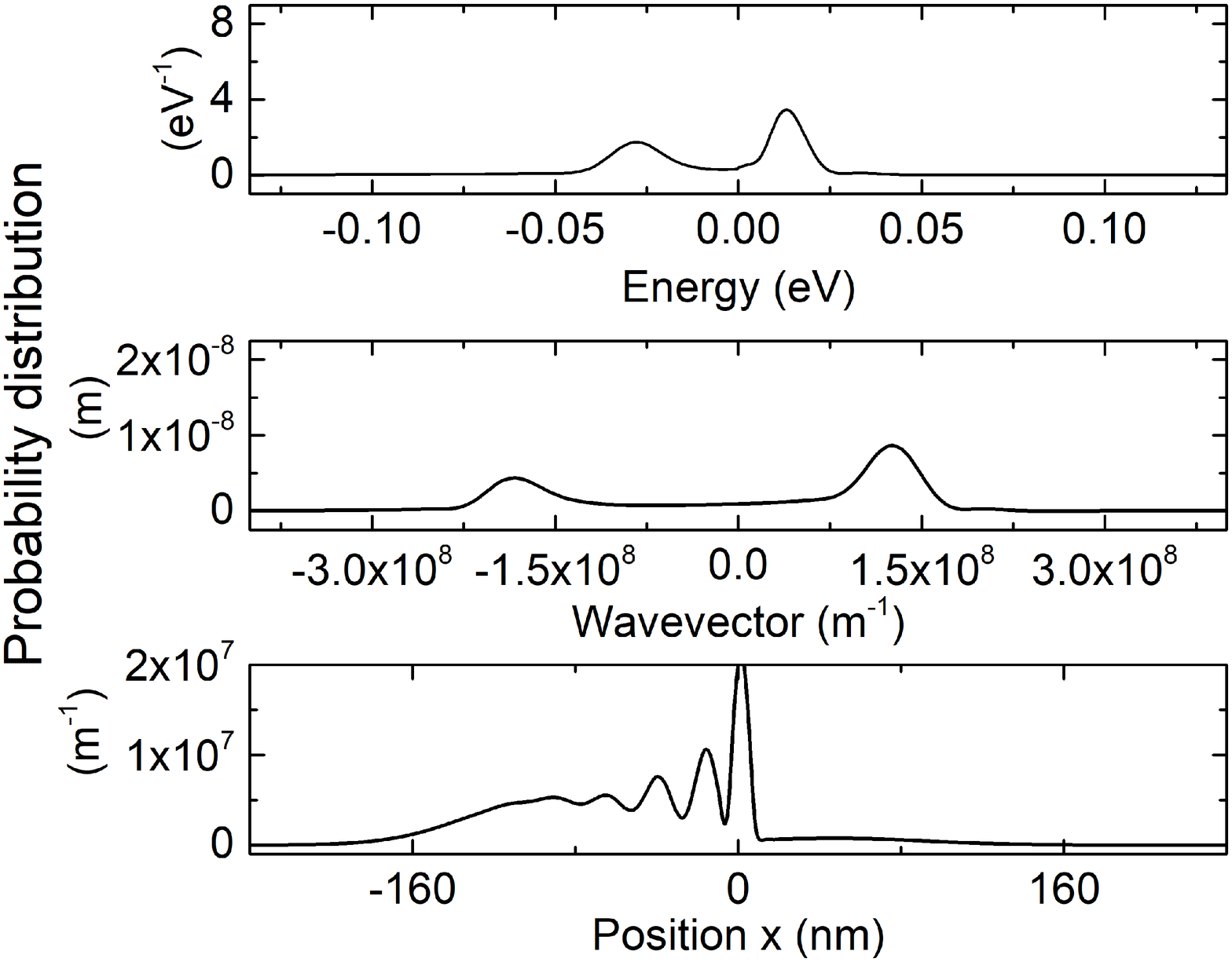}}
	\end{minipage}
	\caption{Comparison of Wigner functions undergoing photon absorption in a double barrier structure (green lines), at different times: (a) before scattering (b) scattered with the energy exchange model and (c) with the momentum exchange model. In (d), (e), (f) are shown projections along the energy (top image), momentum (middle) and position (bottom) axis of the Wigner transform respectively of (a), (b), and (c). Notice the change for the transition from the first resonant level to the second from (d) to (e), while such transition is not present from (d) to (f).}
	\label{comparison_DB_pos}
\end{figure*}

\begin{figure*}[h]
				\hspace{-0.5cm}
		\begin{minipage}{\linewidth}
			{ \includegraphics[scale=0.23]{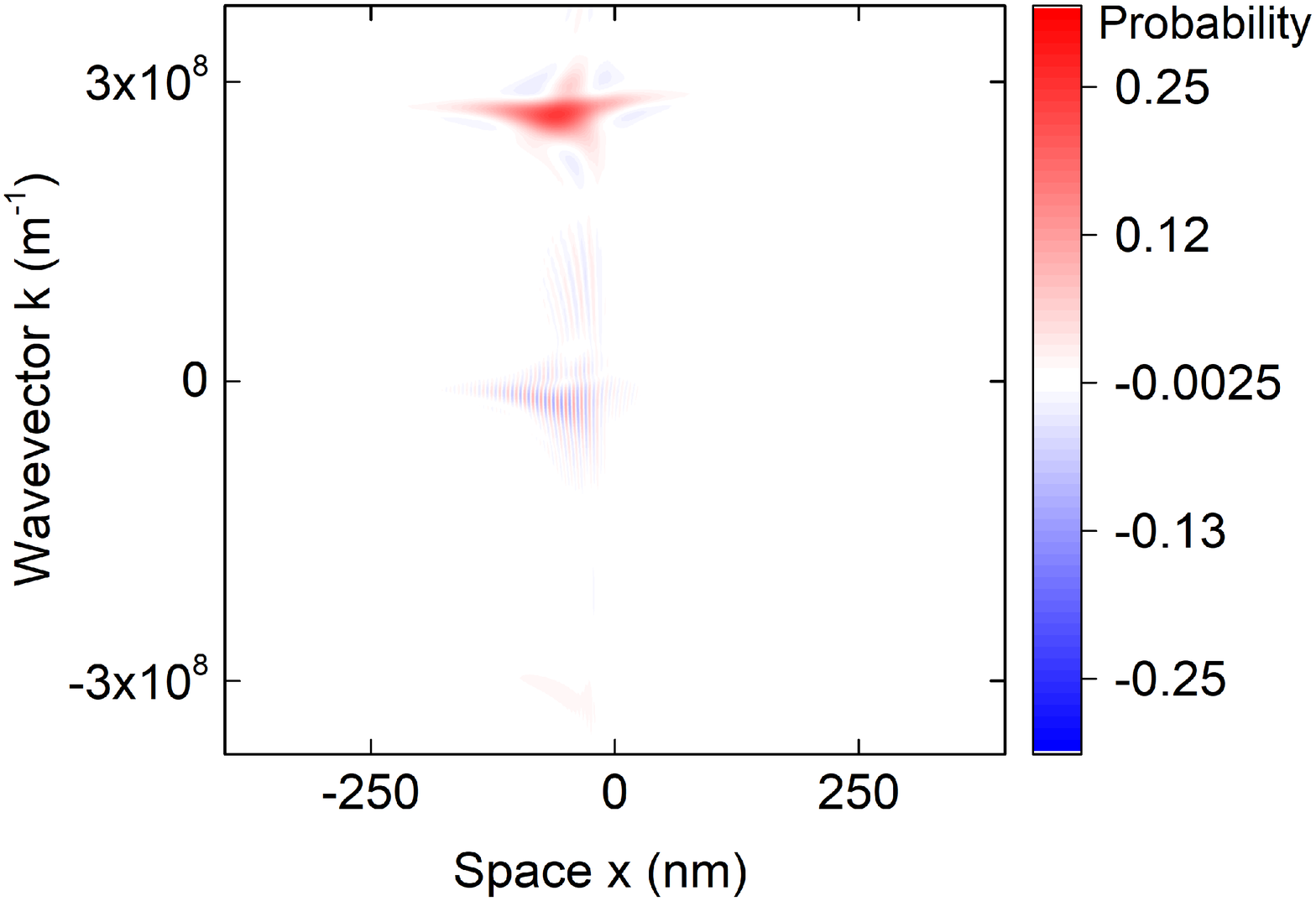}}
			{ \includegraphics[scale=0.23]{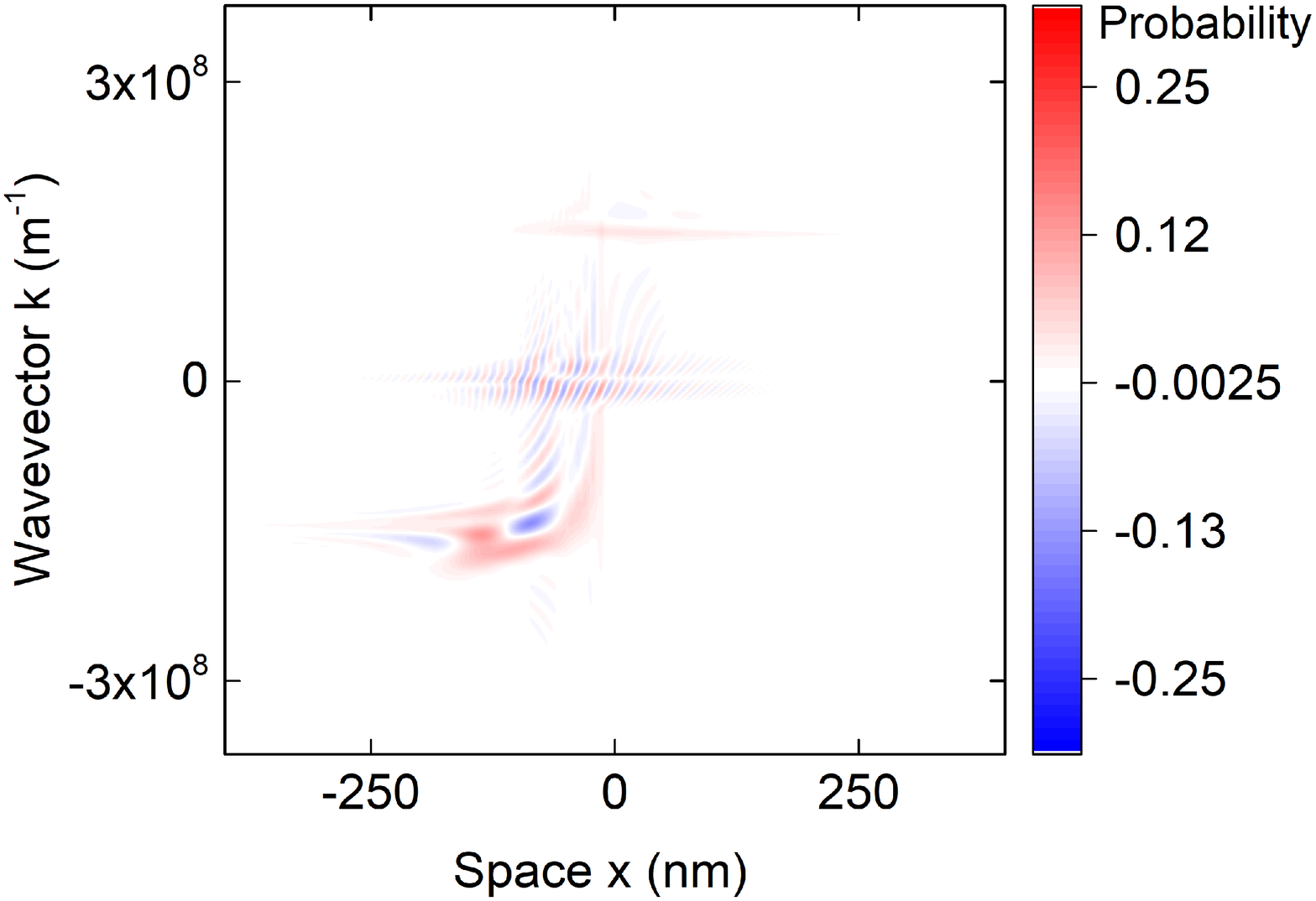}}
			{ \includegraphics[scale=0.23]{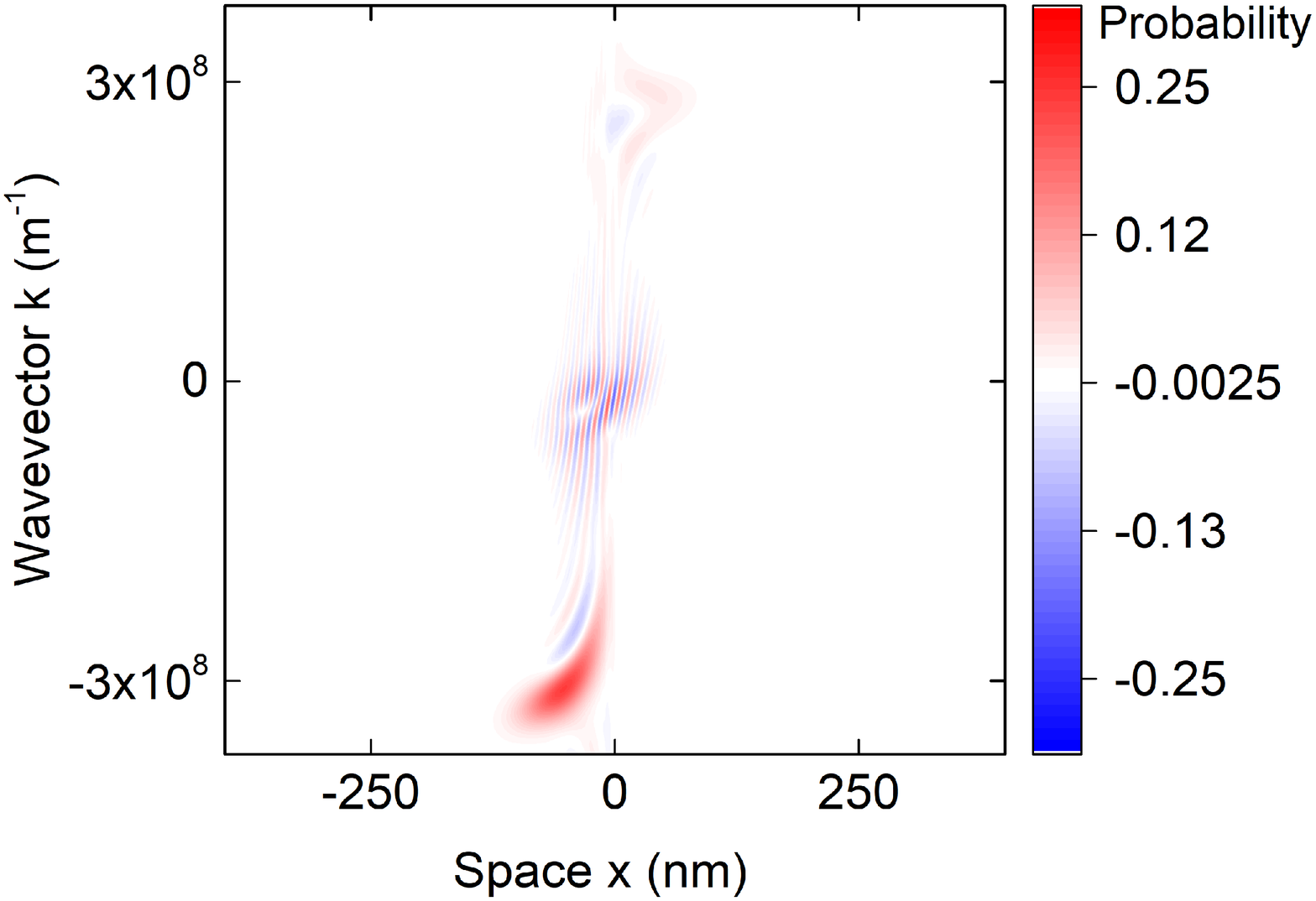}}
			{ \includegraphics[scale=0.23]{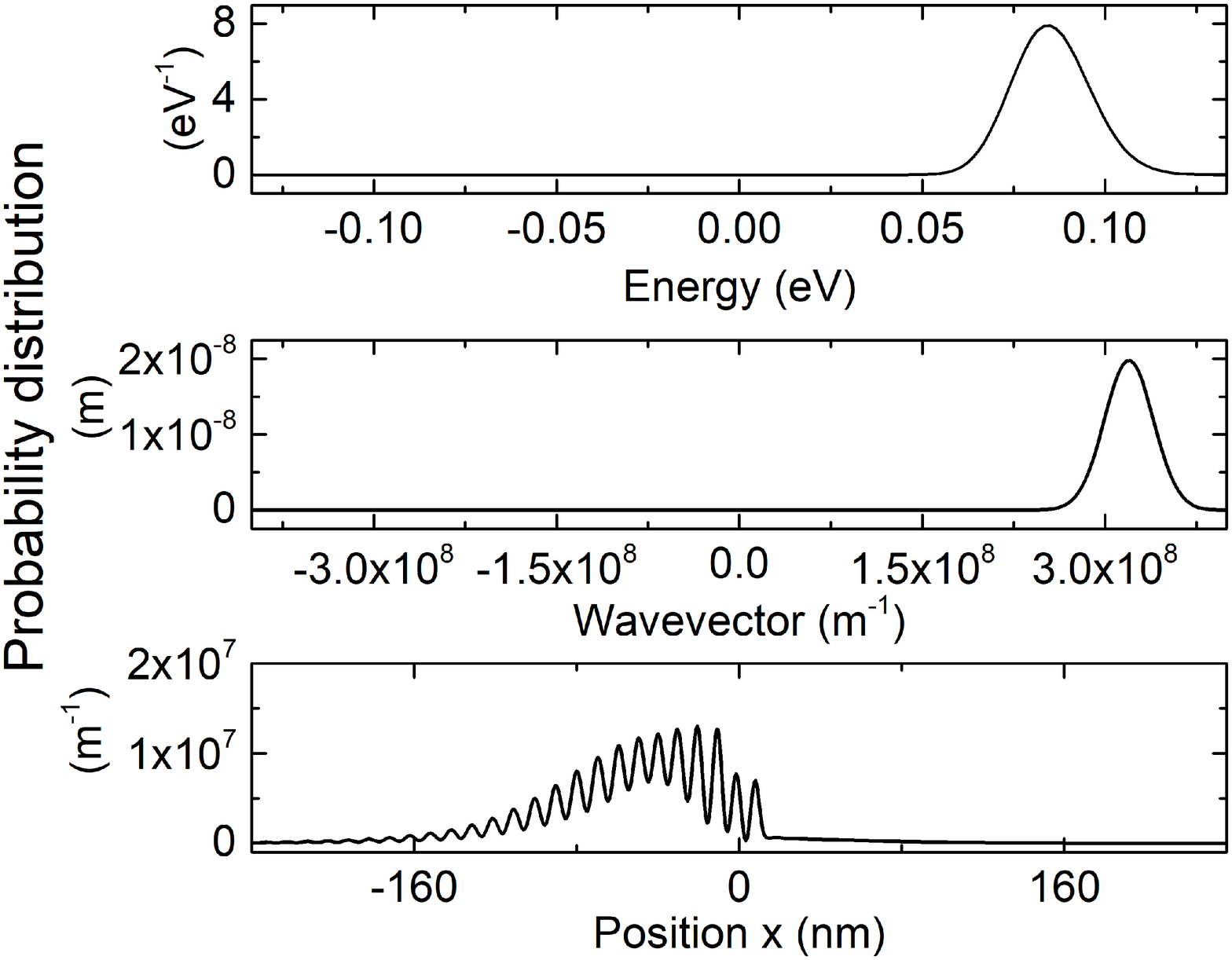}}
			{ \includegraphics[scale=0.23]{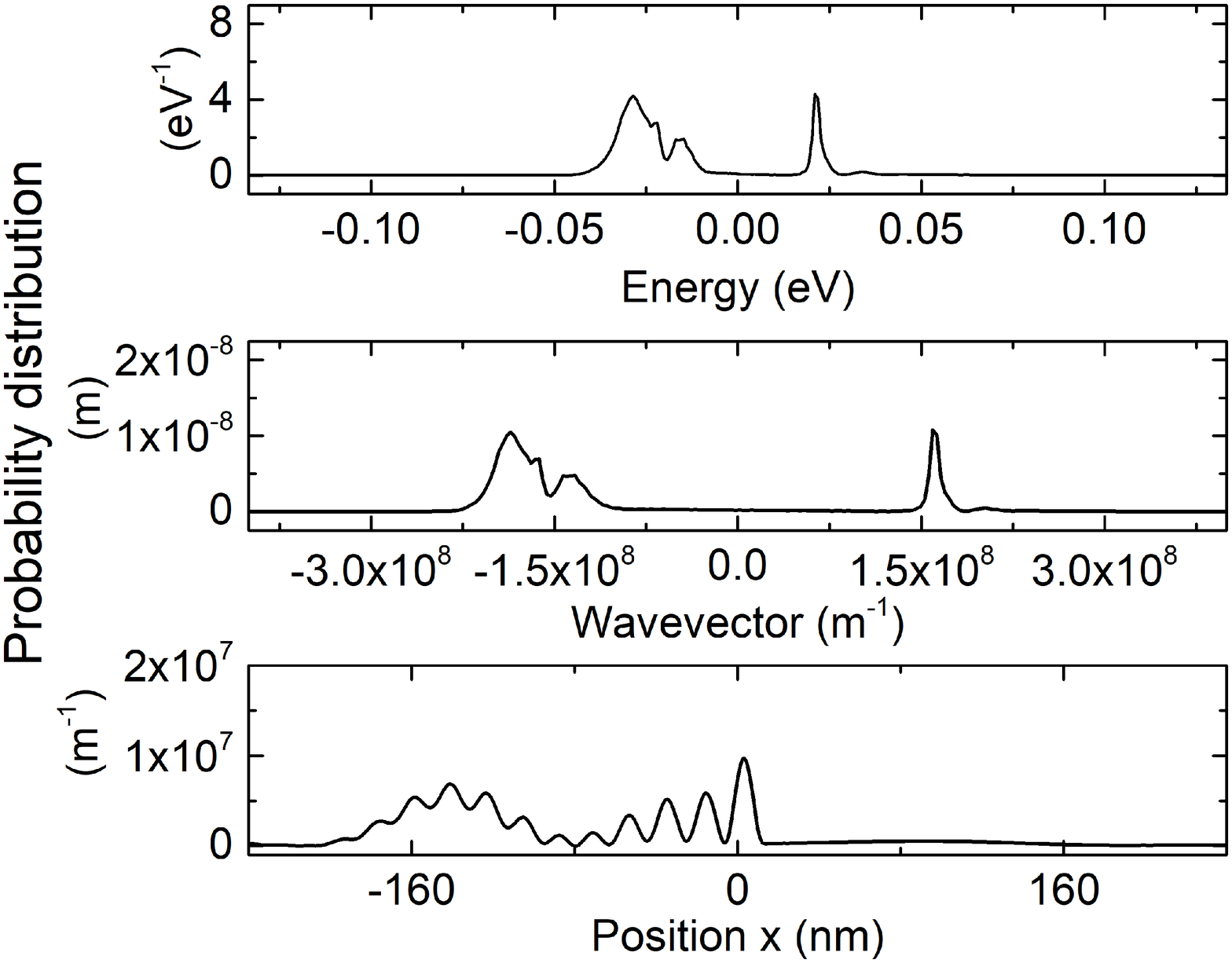}}
			{ \includegraphics[scale=0.23]{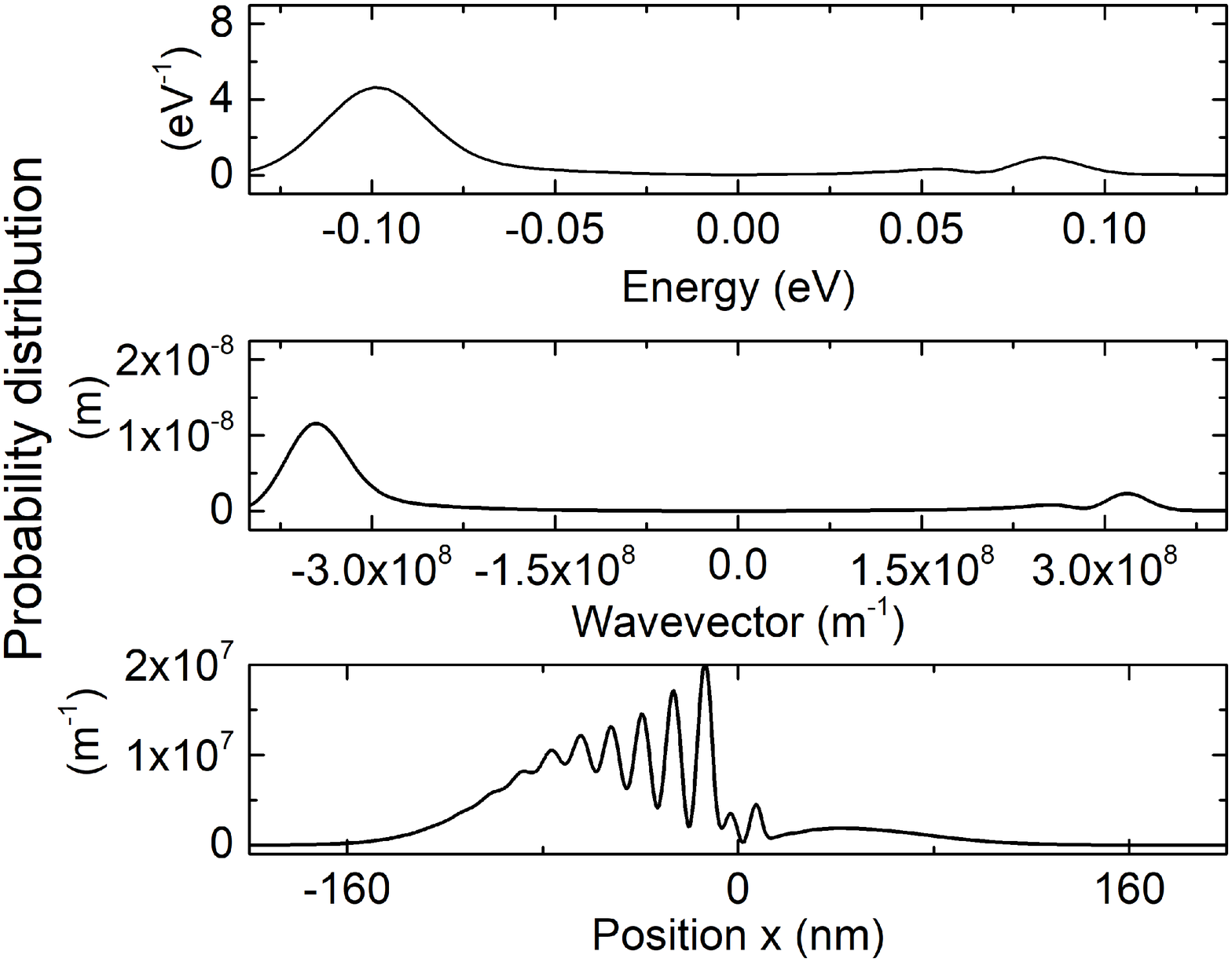}}
		\end{minipage}
		\caption{Comparison of Wigner functions undergoing photon emission in a double barrier structure (green lines), at different times: (a) before scattering (b) scattered with the energy exchange model and (c) with the momentum exchange model. In (d), (e), (f) are shown projections along the energy (top), momentum (middle) and position (bottom) axis of the Wigner transform respectively of (a), (b), and (c). Notice the change for the transition from the second resonant level to the first from (d) to (e), while such transition is not present from (d) to (f).}
		\label{comparison_DB_neg}
	\end{figure*}

In Fig. \ref{comparison_DB_pos}, the electron is first injected at the first resonant level of the double barrier structure and undergoes a photon absorption, thus occupying, after the scattering, the second level of the double barrier structure. Energetically the initial an final energy values are the same as in Fig. \ref{comparison_FP_pos}. This transition is clear in Fig. \ref{comparison_DB_pos}(b) and (e), where the energy exchange model is used, in fact, while in \ref{comparison_DB_pos}(d) there is a single maximum of the probability density in the bottom image, there are two maxima in \ref{comparison_DB_pos}(d). This is a proof of a transition from the first to the second resonant state of the well. However this is not observed in \ref{comparison_DB_pos}(c) and (f), where a momentum exchange model is used. Now, inside the well, the maximum of probability after the scattering belongs still to the first resonant level. This is an unphysical result whose physical reason is that there is no one-to-one correspondence between energy and momentum. Thus, despite having control on the change of the momentum, we do not have control on the change of the energy. In the case of photon emission, shown in Fig. \ref{comparison_DB_neg}, the same physical transition is observed from (a) to (b) (and on their respective projections in (d) and (e), where the two maxima of probability inside the quantum well are transformed into just one maxima later), while the transition between levels is not observed from (a) to (c), where the number of maxima remain equal to two. This is another proof of the difficulties to use a momentum exchange model in case of an arbitrary potential. In other words, in typical device scenarios for nanoscale devices, the energy and momentum operators do not commute so that the eigenstates of the momentum are not eigenstates of the energy and vice-versa. Then, the change in momentum can have an arbitrary translation into change of energy. The dramatic consequence is that the conservation of energy during the electron-photon collision is not guaranteed.

\end{multicols}

\begin{multicols}{2}

%\begin{figure*}[b]

%\end{figure*}

\section{Does Wigner Function satisfy complete positivity and energy conservation?}
\label{s5}

The typical way of reducing the computational burden by including the role of photons and phonons in quantum transport is through a collision model. Such collision model introduces some \text{new} term in the equation of motion of the simulated degrees of freedom. For example, in the Wigner distribution formalism, we can visualize the effect of the collision as a change from the initial (before collision) Wigner function $f_W(x,k,t)$ to the final (after collision) Wigner function $f_W^s(x,k,t)$. But, is there any mandatory requirement that such change of the Wigner function has to satisfy? Or is any possible change physically acceptable?

By construction, the Wigner function without external collisions do satisfy the complete positivity in the sense that $\int^{+\infty}_{-\infty} dk f_W(x,k,t_s)>0$ for all positions $x$. The Wigner function, after being modified by the collision operator, does also have to satisfy the same complete positivity condition:
\begin{eqnarray}
	Tr(\hat \rho(t_s) |x\rangle\langle x|)= \int^{+\infty}_{-\infty} dk f^s_W(x,k,t_s)>0  \;\;\; \forall \;\; x.
	\label{cp}
\end{eqnarray}
Notice that for a discrete Wigner function (used in the computational algorithms), for a grid with $N_x$ position points and $N_k$ wave vector points,  the condition \eqref{cp} imposes $N_x$ equations, each equation involving $N_k$ elements of the estimated discrete $f^s_W(x,k,t_s)$.

On the other hand, during the collision we can assume a controlled change of the energy of the system. To verify such energy conservation, the expectation value of the electron energy within the Wigner approach is defined as:
\begin{equation}
	\langle E(t) \rangle=Tr(\hat{\rho}\hat{H_0})=\int dk \, dx \, f_W(x,k,t) \, h_0(x,k),
	\label{mean_energy}
\end{equation}
where $h_0(x,k)=\int dx' \langle x+\frac{x'}{2}| \hat{H_0} |x-\frac{x'}{2}\rangle e^{-ikx'}$ with $\hat{H_0}$ the Hamiltonian without collisions defined in \eqref{He}. 

The initial energy of the Wigner function is $\langle E(t_s) \rangle=\int^{+\infty}_{-\infty} dx \int^{+\infty}_{-\infty} dk f_W(x,k,t_s) h_0(x,k)$, then after the absorption (or emission) of a photon with energy $E_\gamma$, the final Wigner function has to satisfy:
\begin{eqnarray}
	\langle E^s(t_s) \rangle&=&\langle E(t_s) \rangle \pm E_\gamma=Tr(\hat \rho^s \hat H)\nonumber\\&=&\int^{+\infty}_{-\infty} dx \int^{+\infty}_{-\infty} dk \, f_W^s(x,k,t_s) h_0(x,k).
	\label{ce2}
\end{eqnarray}
The condition \eqref{ce2} implies one additional equation involving  $N_x \times N_k$ elements of the new Wigner function. Together,  conditions \eqref{cp} and \eqref{ce2} do only define $N_x+1$ elements of the total $N_x\times N_k$ elements of the Wigner function after the scattering $f^s_W(x,k,t_s)$. Notice that the Wigner distribution function has $N_x \times N_k$ elements so that these two conditions do not totally determine which is the new Wigner function after scattering. In other words, there are many possible Wigner functions that can satisfy \eqref{cp} and \eqref{ce2}, but there are also many Wigner function that are invalid because they do not satisfy \eqref{cp} and \eqref{ce2}.   

We want to emphasize again that there is no fundamental limitation for developing a successful collisions model from the Wigner function formalism. In other words, it is possible to find $f_W(x,k,t_s)$ and $f^s_W(x,k,t_s)$, before and after the scattering respectively, that satisfy complete positivity and energy conservation. To underline this point, we explain an algorithm that use the collision model developed in Sec. \ref{s32} for wavefunctions. The algorithm explained below will not be easily implementable in a general Wigner transport formalism, but it will clearly confirm that it is possible to develop a successful transition $f_W(x,k,t_s) \to f^s_W(x,k,t_s)$ due to the collision of an electron with a photon (phonon), while satisfying conditions \eqref{cp} and \eqref{ce2}. The algorithm deals with a single electron (as we have done in all this paper), and it has the following three steps:
\begin{itemize}
	\item \textbf{First step:} The Wigner function before the scattering $f_W(x,k,t)$ is translated into a wave function $\psi(x,t_s)$ in the following way:
	\begin{eqnarray}
		&&\psi(x,t_s)=\rho(x,0,t_s)\frac{1}{\psi^*(0,t_s)}\nonumber\\
		&=&\frac{1}{2 \pi \psi^*(0,t_s)}\int\!\!\!\int dx' dk \; \rho \left(y,y',t_s \right)  e^{ik(x-x')} \nonumber\\&=& \frac{1}{\psi^*(0,t_s)}\int f_W\left(\frac{x}{2},k,t_s \right) e^{ikx} dk,
		\label{transform}
	\end{eqnarray}
	where we have used that $\rho(x,0)=\psi^*(0,t)\psi(x,t)$ for a pure state, and $y=(x+x')/2$ and $y'=(x-x')/2$. The complex number $\psi^*(0,t_s)$ is irrelevant here since it can be understood as a normalization constant \cite{pedestrian}.
	
	\item \textbf{Second step:} Once we have $\Psi(x,t_s)$ we can apply the collision algorithm explained in Sec. \ref{s32} to get the transition $\Psi(x,t_s)\to \Psi^s(x,t_s)$ that we know satisfy complete positivity and energy conservation. 
	
	\item \textbf{Third step:} Once we get the pure state after the collision $\Psi^s(x,t_s)$,  we use expression \eqref{Wig_transf} to compute the Wigner distribution function after the collision: $f^s_W(x,k,t_s)$. \\
\end{itemize}

\begin{figure}[h]
	\centering
	{ \includegraphics[scale=0.6]{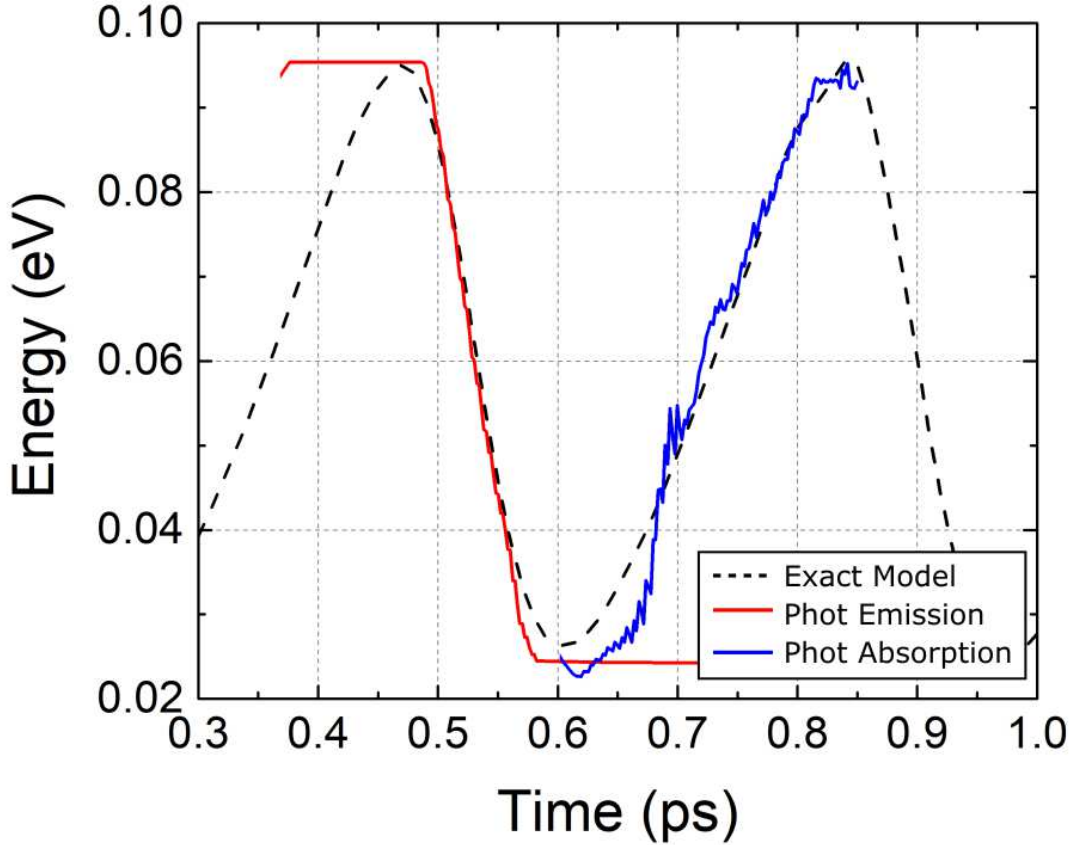}}
	\caption{Comparison between the expectation value of energy  of the system undergoing collision modelled from the exact model (black dashed line) and with the energy exchange algorithm, for photon emission (red line) or photon absorption (blue line). These models are implemented from the Wigner distribution function of a pure state.}
	\label{Energy_comparison}
\end{figure}

Indeed, an example of such Wigner functions, from\\ $f_W(x,k,t_s)$ before collision to $f^s_W(x,k,t_s)$ after collision, is already plotted from Fig. \ref{comparison_DB_pos}(a) to (b), respectively, for photon absorption. Identically, another example is provided from Fig. \ref{comparison_DB_neg}(a) to (b) for photon emission. In Fig. \ref{Energy_comparison} it is shown how such algorithm (based on the electron-photon collision model of Sec. \ref{s32}) reproduce the exact electron-phonon interaction computed from Sec. \ref{s31}. The evolution of the ensemble value of the energy for simulations in Fig. \ref{Exact_DB} (black dashed line), \ref{comparison_DB_pos}(a) and  \ref{comparison_DB_pos}(b) (blue line), \ref{comparison_DB_neg}(a) and \ref{comparison_DB_neg}(b)(red line). The ensemble value $\langle E(t) \rangle$ is computed using expression \eqref{mean_energy} at different times. To better approximate the exact result the change of energy $E_\gamma$ is not done in a single time step of the simulation, but in a time interval identical to the time it takes the exact solution to produce such energy transition (which is related to the frequency of the Rabi oscillations \cite{light}).  
\end{multicols}
\begin{multicols}{2}
The excellent agreement in Fig. \ref{Energy_comparison} was possible because the evolution of such system is treated on the level of the Bohmian conditional wavefunction. Without such knowledge, the task of satisfying conditions \eqref{cp} and \eqref{ce2} becomes very complicated. In a real electron device, more than one electron have to be simultaneously simulated and then expression \eqref{transform} cannot be applied because the density matrix, and its Wigner transform, will hide the knowledge about every single-particle state. We conclude that without the knowledge of the state it seems quite complicated to model a reasonable collision model that satisfies energy conservation. For example,  there is no guarantee at all that the direct implementation of the Boltzmann collision operator in the Wigner distribution can satisfy the above conditions.

\subsection{Are collisions a source of time irreversibility ?}
\label{s51}
At this point, although not directly related with the goal of this paper, it is interesting to discuss if collisions in electron devices are a source of time-irreversibility or not. While typical microscopic laws are time-reversible, an arrow of time appears in macroscopic phenomena. For example, we know that electrons lose energy, in average, when traversing the device (Joule effect). Therefore, we can know if we are looking \emph{forward} or \emph{backward} in time by looking at the macroscopic heating or cooling of the device. But, why microscopic time-reversible laws become time-irreversible at the macroscopic level? What is the role of collisions? 

We have shown in Sec. \ref{s40} that electron-photon interaction can be studied in an exact way as a solution of the many-body Schr\"odinger equation in a closed system including the degrees of freedom of one electron and one photon. Thus, we can conclude that interaction of a single electron with a single photon are a time-reversible phenomena, either for emission or absorption, when described by \eqref{scho} because we know exactly how to go back in time in this exact model, and the process of emission and absorption level are time-reversible at the microscopic. 

In the approximations done in Sec. \ref{s32} and Sec. \ref{s33} we eliminate the explicit simulation of the photon degree of freedom and introduce the effect of the photon into the electron as a collision. The electron alone, without the photon, is an open (non-Markovian) system. But, if we know exactly when the collision with the electron takes place, then we can still consider that approximations done in Sec. \ref{s32} and Sec. \ref{s33} as time-reversible (as it was Eq. \eqref{scho} itself) because we know how to go back in time, and the process of emission and absorption level are time-reversible at the microscopic. 

The conclusions above for time-reversibility of the equation of motion of one  electron interacting with one photon can be straightforwardly generalized to the equation of motion of thousands of electrons interacting with many thousands of photons, as far as we only look at the microscopic information. However, for such many-body system, if instead of looking at the microscopic degrees of freedom, we are interested in a macroscopic equation of motion for the average energy translated from the electrons to the photons and viceversa, the conclusion can be different. If we were able to simulate a many-particle Schr\"odinger equation of such system, including the electron device and the surroundings in an scenario outside of thermodynamic equilibrium,  we will realize that a photon emitted from the electron device will hardly be able to return to the electron device again because the surroundings are larger extension, in space, than the electron device. Thus, inside the electron device there will be more emissions than absorption, in average. In fact, the expected net dissipation of electron energy (Joule effect) will take place. We conclude that the equation of motion for the macroscopic average energy in the electrons is time-irreversible. Notice that the time-irreversibility of the (macroscopic) equation of motion of the average energy of electrons is fully compatible with the time-reversibility of the (microscopic) equation of motion of electrons and photons.

In conclusion, \emph{are collisions a source of time irreversibility?} Not in a microscopic description, but yes in a macroscopic one. Microscopically the collisions provide an equation of motion for the dynamics of electrons and photons that is time-reversible. Macroscopically the collisions ensure that the dissipation of the average energy of the electrons is time-irreversible and satisfy the Joule effect.

We have learn a practical lesson to implement a collision model here. Since the exact solution of thousands of electrons interacting with thousands of photons is not possible, the practical implementation of the scattering model with the approximations done in Sec. \ref{s41} has to satisfy the microscopic time-reversibility of the equation of motion and, simultaneously, the time-irreversibility of the macroscopic average energy. This is typically achieved by randomly selecting the scattering rates with a distributions that satisfies the macroscopic requirement: more emission that absorptions.

\section{Conclusions}
\label{s6}

In this paper we have shown that a physically reasonable collision model, within the Wigner description, has to satisfy the following two mandatory requirements: Complete positivity and energy conservation. 
After presenting the two conditions and studying its implementation for an electron inside a double barrier interacting with a photon, we compare exact results to a collision model based on energy exchange in Sec. \ref{s32} and to a collision model based on momentum exchange in Sec. \ref{s33}. We conclude that, unfortunately, it is very complicated to develop a practical collision model in the Wigner formalism that satisfies both conditions. 

The good conclusion is that there is no fundamental reasons that disqualifies, \textit{a priori}, the possibility of implementing collisions in the Wigner distribution framework. The practical requirement in \eqref{cp} and in \eqref{ce2} provides less restrictions than the degrees of freedom (number of elements) of a (discretized) Wigner distribution function. In fact, for a pure state, we have shown an algorithm that models electron-phonon collisions in the Wigner function formalism with an excellent agreement. The problem for its generalization to realistic quantum transport is that the mentioned algorithm needs the wave function information of each electron inside the device, but such additional information is not available in the typical Wigner function algorithms.   

Finally, we want to mention that  most of the difficulties of the Wigner distribution function to satisfy energy conservation in collisions, are inherited form the difficulties of its father description, the density matrix, to tackle the properties of individual particles. At the end of the day, these problems are just a manifestation of the orthodox statements that negates any microscopic properties for particles, unless such microscopic properties are being measured explicitly \cite{Proceddings}. But, a collision is not a measurement, so that the orthodox theory forbids to access to the microscopic information on what has happened to each electron during the collision. For these reasons, we argue that collisions models based on the (Bohmian) conditional wave function are very promising because they have the ability to describe the microscopic properties of individual particles and satisfy conditions \eqref{cp} and \eqref{ce2} in a very \emph{natural} way \cite{Proceddings,PRLxavier,Entropy,BookXavier,BITLLES1,Albareda,Marian}.

%\end{paracol}
%%%%%%%%%%%%%%%%%%%%%%%%%%%%%%%%%%%%%%%%%%
\vspace{6pt}

\paragraph{Contributions}
Conceptualization: Matteo Villani, Xavier Oriols; Methodology: Matteo Villani, Xavier Oriols; Formal analysis and investigation: Matteo Villani, Xavier Oriols; Writing - original draft preparation: Matteo Villani, Xavier Oriols; Writing - review and editing: Matteo Villani, Xavier Oriols; Funding acquisition: Xavier Oriols; Resources: Xavier Oriols; Supervision: Xavier Oriols. All authors read and approved the final manuscript.

\paragraph{Funding}
This project has received funding from the European Union's Horizon 2020 research and innovation programme under the Maria Skłodowska-Curie grant agreement No. 765426 (TeraApps). This research was also funded by Spain's Ministerio de Ciencia, Innovaci\'on y Universidades under Grant No. RTI2018-097876-B-C21 (MCIU/AEI/FEDER, UE), the "Generalitat de Catalunya" and FEDER for the project 001-P-001644 (QUANTUMCAT).

\paragraph{Conflicts of interest}
The authors have no conflicts of interest to declare that are relevant to the content of this article. The funders had no role in the design of the study; in the collection, analysis, or interpretation of data; in the writing of the manuscript, or in the decision to publish the results.

\paragraph{Compliance with Ethical Standards}
Not applicable.

\paragraph{Availability of data and material} 
The datasets generated during and/or analysed during the current study are available from the corresponding author on reasonable request.
\end{multicols}


\begin{thebibliography}{}
	
	\bibitem{open} %Summary
	Breuer, H. P., Petruccione, F. \textit{Theory of Open Quantum Systems}, Oxford University Press, Oxford, 2002; pp. 5-23.
	
	\bibitem{vega} %Review non-markovianity
	Vega, I., Alonso, D.: Dynamics of non-Markovian open quantum systems. Rev. Mod. Phys. 89, 015001 (2017) https://doi.org/10.1103/RevModPhys.89.015001
	
	\bibitem{Klimeck}
	Klimeck, G.: Single and multiband modeling of quantum electron transport through layered semiconductor devices. J. App. Phys. 81, 7845 (1997). https://doi.org/10.1063/1.365394
	
	\bibitem{Klimeck2}
	Klimeck, G., Ahmed, S.S., Bae, H., Kharche, N., Clark, S., Haley, B., Lee, S., Naumov, M.,  Ryu, H., Saied, F., Prada, M.,  Korkusinski, M., Boykin, T.B.: Atomistic Simulation of Realistically Sized Nanodevices Using NEMO 3-D Part I: Models and Benchmarks. IEEE Trans. Electron Devices 54, 9, 2079-2089 (2007). https://doi.org/10.1109/TED.2007.902879
	
	\bibitem{Green}
	Schmidt, A., Cheng, B., DaLuz, M.: Green function approach for general quantum graphs. J. Phys. A, 36, 42 (2003). https://doi.org/10.1088/0305-4470/36/42/L01
	
	\bibitem{Rossi1}
	Rossi, F. The Density-Matrix Approach Theory of Semiconductor Quantum Devices. \textit{NanoScience and Technology} Springer, Berlin, Heidelberg, 2010, pp. 89-130.
	
	\bibitem{Rossi2}
	Iotti, C., Ciancio, E., Rossi, F.: Quantum transport theory for semiconductor nanostructures: A density-matrix formulation. Phys. Rev. B, 72, 125347 (2005). https://doi.org/10.1103/PhysRevB.72.125347
	
	\bibitem{Ferry} %intro Wigner
	Weinbub, J.; Ferry, D. K.: Recent advances in Wigner function approaches. Appl. Phys. Rev. 5, 041104 (2018). https://doi.org/10.1063/1.5046663
	
	\bibitem{Frensley}
	Frensley, W.; Wigner-Function Model of Resonant-Tunneling Semiconductor Device. Phys. Rev. B. 36, 3, 1570–1580 (1987).
	
	\bibitem{Dollfus1}  %intro Wigner
	Querlioz, D., Huu-Nha Nguyen, Saint-Martin, J.,  Bournel,  A., Galdin-Retailleau, S., Dollfus, P.: Wigner-Boltzmann Monte Carlo approach to nanodevice simulation: from quantum to semiclassical transport. J. Comput. Electron. 8, 324–335 (2009). https://doi.org/10.1007/s10825-009-0281-3
	
	\bibitem{Dollfus2} %intro Wigner
	Nedjalkov, M., Querlioz, D., Dollfus, P., Kosina, H.: Wigner Function Approach. Nano-Electronic	Devices. Springer, (2011). pp. 289–358.
	
	\bibitem{Nedjalkov} %intro Wigner
	Nedjalkov, M., Selberherr, S., Ferry, D.K., Vasileska, D.,
	Dollfus, P., Querlioz, D., Dimov, I., Schwaha, P.: Physical scales in the Wigner-Boltzmann equation. Ann. Phys.
	328, 220-237 (2013). https://doi.org/10.1016/j.aop.2012.10.001
	
	\bibitem{Jonasson}Jonasson, O., Knezevic, I. : Coulomb-driven terahertzfrequency 	intrinsic current oscillations in a double-barrier tunneling structure. Phys. Rev. B 90, 165415 (2014). https://doi.org/10.1103/PhysRevB.90.165415
	
	\bibitem{Polkovnikov} Polkovnikov, A.: Phase space representation of quantum
	dynamics. Ann. Phys. 325, 1790-1852 (2010). https://doi.org/10.1016/j.aop.2010.02.006
	
	\bibitem{Fischetti1}
	Vyas, P., Van de Put, M., Fischetti, M.: Master-Equation Study of Quantum Transport in Realistic Semiconductor Devices Including Electron-Phonon and Surface-Roughness Scattering. Phys. Rev. Applie. 13, 014067, (2020). https://doi.org/10.1103/PhysRevApplied.13.014067
	
	\bibitem{Fischetti2}
	Fischetti, M.: Theory of electron transport in small semiconductor devices using the Pauli master equation, J. Appl. Phys. 83, 270, (1998). https://doi.org/10.1063/1.367149
	
	\bibitem{Cummings}
	Fan, Z., Garcia, J., Cummings, A., Barrios-Vargas, J., Panhans, M., Harju, A., Ortmann, F., Roche, S.: Linear scaling quantum transport methodologies. Physics Reports 903, 1 (2020). http://dx.doi.org/10.1016/j.physrep.2020.12.001
	
	
	\bibitem{Bohm_original}
	Bohm, D.: A Suggested Interpretation of the Quantum Theory in Terms of "Hidden" Variables. Phys. Rev. 85, 166 (1952).
	
	\bibitem{Bohm1}
	D\"urr, D., Teufel, S.: Bohmian Mechanics: The Physics and Mathematics of Quantum Theory. Springer, Berlin (2009).
	
	\bibitem{Proceddings} 
	Oriols, X., Ferry, D. K.: Why engineers are right to avoid the quantum reality offered by the orthodox theory?
	Proc. IEEE. 109, 955
	(2021). https://doi.org/10.1109/JPROC.2021.3067110
	
	\bibitem{PRLxavier}
	Oriols, X.: Quantum-Trajectory Approach to Time-Dependent Transport in Mesoscopic Systems with Electron-Electron Interactions. Phys. Rev. Lett. 98, 6, 066803 (2007). https://doi.org/10.1103/PhysRevLett.98.066803
	
	\bibitem{Entropy}
	Villani, M., Destefani, C., Albareda, G., Cartoixá, X., Oriols, X.: Scattering in Terms of Bohmian Conditional Wave Functions for Scenarios with Non-Commuting Energy and Momentum Operators, Entropy, 23(4), 408 (2021). https://doi.org/10.3390/e23040408
	
	\bibitem{Wigner}
	Wigner, E.P.: On the quantum correction for thermodynamic equilibrium. Phys. Rev. , 40, 5, 749–759 (1932).
	
	\bibitem{Zhen_true} Colom\'es, E., Zhan, Z., Oriols, X.: Comparing Wigner, Husimi and Bohmian distributions: Which one is a true probability distribution in phase space? J. Comput. Electron. Special issue: Wigner function. 14, 894 (2015) https://doi.org/10.1007/s10825-015-0737-6
	
	\bibitem{Zhen} Zhan, Z., E. Colom´es, E., Oriols, X.: Unphysical features in the application of the Boltzmann collision operator in the time dependent modelling of quantum
	transport. J. Comput. Electron. 15, 1206 (2016) https://doi.org/10.1007/s10825-016-0875-5
	
	%%%%%%%%%%%%%%%%%%%%%%%%%%%%%%%
	\bibitem{Rossi3}
	Rossi, F., Kuhn, T.: Theory of ultrafast phenomena in photoexcited semiconductors. Rev. Mod. Phys. 74, 895 (2002) https://doi.org/10.1103/RevModPhys.74.895
	
	\bibitem{Nedjalkov2}
	Nedjalkov, M., Vasileska, D., Ferry, D. K., Jacoboni, C., Ringhofer, C., Dimov, I., Palankovski, V.: Wigner transport models of the electron-phonon kinetics in quantum wires. Phys. Rev. B. 74, 035311 (2006) https://doi.org/10.1103/PhysRevB.74.035311
	
	\bibitem{Gurov}
	Gurov, T.V., Nedjalkov, M., Whitlock, P.A., Kosina, H., Selberherr, S.: Femtosecond relaxation of hot electrons by phonon emission in presence of electric field. Phys. B: Condensed Matter.  314(1-4), 301 (2002) https://doi.org/10.1016/S0921-4526(01)01417-X
	
	\bibitem{deva}
	Pandey D., Colom\'es, E., Albareda G., Oriols X.: Stochastic Schrodinger Equations and Conditional States: A General Non-Markovian Quantum Electron Transport Simulator for THz Electronics, Entropy. 21(12), 1148 (2019) https://doi.org/10.3390/e21121148
	
	\bibitem{wiseman1}
	Gambetta, J.; Wiseman, H.: Non-Markovian stochastic Schrödinger equations: Generalization to real-valued noise using quantum-measurement theory. Phys. Rev. A. 66, 012108 (2002) https://doi.org/10.1103/PhysRevA.66.012108
	
	\bibitem{wiseman2}
	Gambetta, J.; Wiseman, H.: Interpretation of non-Markovian stochastic Schrödinger equations as a hidden-variable theory. Phys. Rev. A.  68, 062104 (2003) https://doi.org/10.1103/PhysRevA.68.062104
	
	\bibitem{light} Grynberg G., Aspect A., Fabre C.: Introduction to Quantum Optics: From the Semi-classical Approach to Quantized Light.
	Cambridge University Press, New York (2010). https://doi.org/10.1017/CBO9780511778261
	
	\bibitem{Enrique1}
	Colom\'es, E., Zhan, Z., Marian, D., Oriols, X.: Quantum dissipation with conditional wave functions: Application to the realistic simulation of nanoscale electron devices. Phys. Rev. B. 96, 7, 075135 (2017). https://doi.org/10.1103/PhysRevB.96.075135
	
	\bibitem{pedestrian} Case W.: Wigner functions and Weyl transforms for pedestrians. Am. J. of Phys. 76, 10, 937-946 (2008). https://doi.org/10.1119/1.2957889 
	
	\bibitem{BookXavier}
	Oriols, X., Mompart, J. \textit{Applied Bohmian Mechanics: From Nanoscale Systems to Cosmology}, 2nd ed., Jenny Stanford Publishing: Singapore, 2019.
	
	\bibitem{BITLLES1}
	Bohmian Interacting Transport in non-equiLibrium eLEctronic Structures (BITLLES).\\
	Simulator available online: http://europe.uab.es/bitlles.
	
	\bibitem{Albareda}
	Albareda, G., L\'{o}pez, H., Cartoix\`{a}, X., ~Su\~{n}\'{e}, J., Oriols, X.: 
	Time-dependent boundary conditions with lead-sample Coulomb correlations: Application to classical and quantum nanoscale electron device simulators. Phys. Rev. B. 82, 085301 (2010) https://doi.org/10.1103/PhysRevB.82.085301
	
	\bibitem{Marian}
	Marian, D., Colom\'es, E., Oriols, X.: Quantum noise from a Bohmian perspective: fundamental understanding and practical computation in electron devices. J. Phys. Condens. Matter. 27, 245302 (2015). https://doi.org/10.1007/s10825-015-0672-6
	
\end{thebibliography}
\end{document}